
\input harvmac

\def\ie{{\it i.e.}}
\def\eg{{\it e.g.}}
\def\bdi{\overline{\cal D}}

\def\CP{{\cal P}}
\def\CM{{\cal M}}

\def\soint{{\textstyle{\oint}}}
\def\ttstar{$tt^\ast$}

\def\Vsl#1{\,\raise.15ex\hbox{/}\mkern-13.5mu #1}
\def\tra#1{{\rm Tr}\left[(-1)^F #1\right]}

\def\BZ{{\bf Z}}

\def\BP{{\bf P}}

\lref\ising{L. Onsager, Phys. Rev. 65 (1944) 117\semi T.T. Wu and
B.M. McCoy,
    {\it The two dimensional Ising Model}, Harvard University Press,
Cambridge,
    Mass. 1973\semi B. M. McCoy and T.T. Wu, Phys. Rev, Lett, 45
(1980) 675;
    B.M. McCoy, J.H.H. Perk and T.T. Wu, Phys. Rev. Lett. 46 (1981)
757.}
\lref\japanese{M. Sato, T. Miwa and M. Jimbo, Publ. R.I.M.S. 14
(1978) 223' 15
    (1979) 201; 577; 871; 16 (1980) 531; 17 (1981) 137\semi M. Jimbo
and T.
Miwa,
    {\it Aspects of holonomic quantum fields}, Lecture Notes in Phys.
vol.126,
    Springer 1980 p.429-491.\semi M. Jimbo and T, Miwa,
    {\it Integrable Systems
    and Infinite Dimensional Lie Algebras}, in {\it Integrable
Systems in
    Statistical Mechanics}, Ed. G.M. D'Ariano, A. Montorsi, M.G.
Rasetti, World
    Scientific, Singapore, 1988\semi  M. Jimbo, Proceedings of
Symposia in Pure
    Mathematics, 49 (1989) 379.}
\lref\hodge{P. Griffiths {\it Topics in Transcendental Algebraic
Geometry},
    Annals of Mathematical Studies 106, Princeton University Press,
    Princeton, 1984.}
\lref\disorder{L.P. Kadanoff and H. Ceva, Phys. Rev. B3 (1971)
3918\semi
    E.C. Marino and J.A. Swieca, Nucl. Phys. B170 [FS1] (1980) 175.}
\lref\hitchen{N. Hitchin, Adv. in Math. 14 (1974) 1.}
\lref\WKD{E. Witten, Nucl. Phys. B258 (1985) 75.}
\lref\compact{S. Cecotti,L. Girardello and A. Pasquinucci, Int. J.
Mod. Phys.
    A6 (1991) 2427.}
\lref\residueapp{S. Cecotti, Int. J. Mod. Phys. A6 (1991) 1749.}
\lref\deligne{P. Deligne, {\it Equations differentielles a points
singuliers
    reguliers}, Lectures Notes in Math. 163, Springer--Verlag 1970.}
\lref\topatop{S.Cecotti and C. Vafa, Nucl. Phys. B367 (1991) 359}
\lref\TFT{E. Witten, Comm. Math. Phys. 118 (1988) 411\semi
    E. Witten, Nucl. Phys. B340 (1990) 281\semi
    T. Eguchi and S.K. Yang, Mod. Phys. Lett. A5 (1990) 1693\semi
    C. Vafa, Mod. Phys. Lett. A6 (1991) 337\semi R. Dijkgraaf, E.
Verlinde, and
    H. Verlinde, Nucl. Phys. B352 (1991) 59.}
\lref\newindex{S. Cecotti, P. Fendley, K. Intriligator and C. Vafa,
{\it A New
    Supersymmetric Index}, preprint Harvard HUTP-92/A021, SISSA
68/92/EP,
    BUHEP-92-14, (1992). }
\lref\sigmamodels{S. Cecotti and C. Vafa, Phys. Rev. Lett. 68 (1992)
903\semi
    S. Cecotti and C. Vafa, Mod. Phys. Lett. A7 (1992) 1715. }
\lref\polymers{P. Fendley and H. Saleur, Boston and Yale preprint
    BUHEP-92-15, YCTP-P13-1992.}
\lref\veryold{S. Cecotti, Int. J. Mod. Phys. A6 (1991) 1749\semi
    S. Cecotti, Nucl. Phys. B355 (1991) 755.}
\lref\principles{P.A.M. Dirac, {\it The Principles of Quantum
Mechanics},
    4th Edition, Oxford University Press, Oxford 1958.}
\lref\chiralring{W. Lerche, C. Vafa and N. Warner, Nucl. Phys. B324
(1989)
427.}
\lref\zamolo{A.B. Zamolodchikov, JETP Lett. 43 (1986) 730.}
\lref\flatcoordinates{R. Dijkgraaf, E. Verlinde and H. Verlinde,
    Nucl. Phys. B352 (1991) 59\semi
    B. Blok and A. Var\v cenko, {\it Topological Conformal Field
Theories
    and the Flat Coordinates}, preprint IASSNS--HEP--91/5, January
    1991\semi M. Saito Publ. RIMS, Kyoto Univ. 19 (1983) 1231\semi
    M. Saito, Ann. Inst. Fourier (Grenoble) 39 (1989) 27.}
\lref\FMVW{P. Fendley, S.D. Mathur, C. Vafa and N.P. Warner, Phys.
Lett. B243
    (1990) 257.}
\lref\cancoordinates{B. Dubrovin, {\it Integrable systems in
topological field
    theory} preprint Napoli INFN-NA-IV-91/26, DSF-T-91/26 (1991).}
\lref\solttstar{B. Dubrovin, {\it Geometry and integrability of
topological
    anti--topological fusion}, Napoli preprint INFN-8/92-DSF.}
\lref\landgins{E. Martinec, Phys. Lett. 217B (1989) 431\semi C. Vafa
and N.P.
    Warner, Phys. Lett. 43 (1989) 730.}
\lref\axioms{K. Osterwalder and R. Schrader, Comm. Math. Phys. 31
(1973)
    83\semi K. Osterwalder and R. Schrader, Comm. Math. Phys. 42
(1975)
    281\semi B. Simon, {\it  The $P(\phi)_2$ Euclidean (Quantum)
Field Theory},
    Princeton University Press 1974.}
\lref\specgeom{S. Ferrara and A. Strominger, {\it N=2 spacetime
supersymmetry
    and Calabi--Yau moduli space},
    presented at Texas A \&\ M University, String'
    89 Workshop\semi S. Cecotti, Commun. Math. Phys. 131 (1990)
517\semi A.
    Strominger, Commun. Math. Phys. 133 (1990) 163\semi P. Candelas
and X.C. de
    la Ossa, {\it Moduli Space of Calabi--Yau Manifolds}, University
of Texas
    Report, UTTG--07--90\semi R. D'Auria, L. Castellani and S.
Ferrara, Class.
    Quant. Grav. 1 (1990) 1767.}
\lref\isingFF{J. Cardy and G. Mussardo, Nucl. Phys. B340 (1990)
387\semi
    V.P. Yurov and Al.B. Zamolodchikov, Int. Mod. Phys.A6 (1991)
3419.}
\lref\piiimath{B. McCoy, J. Tracy, and T.T. Wu, J. Math. Phys. 18
(1977) 1058
    \semi A.R. Its and V.Yu. Novokshenov, {\it The Isomonodromic
    Deformation Method in the Theory of Painlev\'e Equations},
Lectures
    Notes in Mathematics 1191, Springer 1986.}
\lref\raysinger{D.B. Ray and I.M. Singer, Adv. Math. 7 (1971)
145\semi
    D.B. Ray and I.M. Singer, Ann. Math. 98 (1973) 154.}
\lref\susyQM{S. Elitzur and A. Schwimmer, Nucl. Phys. B226 (1983)
109\semi
    M. Claudson and M.B. Halpern, Nucl. Phys. B250 (1985) 689.}
\lref\asforms{E. Witten, J. Diff. Geom. 17 (1982) 661.}
\lref\pathint{E.C. Marino, B. Schroer and J.A. Swieca, Nucl. Phys
B200 (1982)
    473.}
\lref\marino{E.C. Marino, Nucl. Phys. B217 (1983) 413\semi E.C.
Marino, Nucl.
    Phys. B230 (1984) 149.}
\lref\french{O. Babelon, {\it From form factors to correlation
functions: the
    Ising model}, preprint Saclay SPhT-92-062; LPTHE-92-20.}
\lref\highermod{B. Dubrovin, {\it Differential Geometry of Moduli
Spaces
    and its Application to Soliton Equation and to Topological
Conformal
    Field Theory}, Preprint 117 of Scuola Normale Superiore, Pisa,
    November 1991.}
\lref\riemann{H. Flaschka and A.C. Newell, Commun. Math. Phys. 76
(1980) 67.}
\lref\oneloop{L.J. Dixon, V.S. Kaplunovsky and J. Louis, Nucl. Phys.
B355
(1991)
    649\semi I. Antoniadis, E. Gava and K.S. Narain, preprints
IC/92/50 and
    IC/92/51
    \semi S.Ferrara, C.Kounnas, D.L\"ust and F.Zwirner,
    preprint CERN-TH.6090/91.}
\lref\bill{W. Leaf--Herrmann, Harvard preprint HUTP-91-A061, and to
appear.}
\lref\witindex{E. Witten, Nucl. Phys. B202 (1982).}
\lref\markoff{A.A. Markoff, Math. Ann. 15 (1879) 381\semi
    A. Hurwitz, Archiv. der Math. und Phys. 3 (14) (1907) 185\semi
    L.J. Mordell, J. Lond. Math. Soc. 28 (1953) 500\semi
    H. Schwartz and H.T. Muhly, J. Lond. Math. Soc. 32 (1957) 379.}
\lref\mordell{L.J. Mordell, {\it Diophantine Equations},
    Academic Press, London 1969.}
\lref\lambdamat{F.R. Gantmacher, {\it The Theory of Matrices},
    Chelsea, 1960.}
\lref\monlemma{F. Lazzeri, Some remarks on the Picard--Lefschetz
monodromy,
     in {\it Quelques journ\'ees singuli\`eres}, Centre de
Mathematique de
     l'Ecole Polytechnique, Paris 1974.}
\lref\krona{L. Kronecker, {\it Zwei S\"atze \"uber Gleichungen mit
     ganzzahligen Coefficients}, Crelle 1857, Oeuvres 105.}
\lref\cotech{F.M. Goodman, P. de la Harpe, V.F.R. Jones, {\it Coxeter
     Graphs and Tower of Algebras}, Mathematical Sciences Research
     Institute Publications 14, Springer--Verlag, 1989.}
\lref\gross{B.H. Gross, Inv. Math. 45 (1978) 193.}
\lref\weilA{A. Weil, {\it Introdution a les Vari\'et\'es
K\"ahl\'eriennes},
     Hermann, Paris, 1958.}
\lref\griHarris{P. Griffiths and J. Harris, {\it Principles of
Algebraic
     Geometry}, Wiley--Interscience, New York, 1978.}
\lref\singT{V.I. Arnold, S.M. Gusein--Zade and A.N. Var\v cenko,
     {\it Singularities of Differentiable Maps}, Vol.II,
     Birkh\"ausser, Boston 1988.}
\lref\strominger{A. Strominger, Commun. Math. Phys. 133 (1990) 163.}
\lref\pathrep{S. Cecotti and L. Girardello, Phys. Lett. B110 (1982)
39.}
\lref\acampo{N. A'Campo, Indag. Math. 76 (1973) 113.}
\lref\cyclopoly{K. Ireland and M. Rosen, {\it A Classical
Introduction
     to Modern Number Theory}, (Springer--Verlag, Berlin, 1982).}
\Title{ HUTP-92/A044, SISSA-167/92/EP }
{\vbox{\centerline{Ising Model and $N=2$ Supersymmetric Theories } }}

\bigskip
\bigskip

\centerline{Sergio Cecotti}
\medskip\centerline{International School for Advanced Studies,
SISSA-ISAS}
\centerline{Trieste and I.N.F.N., sez. di Trieste}
\centerline{Trieste, Italy}
\bigskip\bigskip

\centerline{Cumrun Vafa}
\medskip\centerline{Lyman Laboratory of Physics}
\centerline{Harvard University}
\centerline{Cambridge, MA 02138, USA}

\vskip .3in

We establish a direct link between massive Ising model and arbitrary
massive $N=2$ supersymmetric QFT's in two dimensions.  This explains
why the
equations which appear in the computation of
 spin-correlations in the non-critical Ising model are the same as
those
describing the geometry of vacua
in $N=2$ theories.  The tau-function appearing in the Ising
model (i.e., the spin correlation function)
 is reinterpreted in the $N=2$ context as a new `index'.
In special cases this new index is related to Ray-Singer analytic
torsion, and can be viewed as a generalization of that
to loop space of K\"ahler manifolds.

\Date{9/92}

\newsec{Introduction}
Two dimensional Ising model was among the first integrable models to
be studied
\ising .
It is a very rich theory and has played a prominent role in the
development of integrable hierarchies.  In particular the notion of
a tau function arose as a result of studying this theory.  This line
of development arose by the realization \ising\ that
for non-critical Ising model, which is equivalent to free
massive fermions,  spin correlations
can be computed by relating them to solutions of certain differential
equations.  For example, for the
two point function the corresponding equation turns out to be a
special case
of Painleve III.    Formalizing these ideas and generalizing them
led the Japanese school \japanese\ to a set of differential equations
in $n $ variables,
whose solutions could be used to construct $n$ point spin
correlation function for massive free fermion theory;
the first example of a {\it tau} function.  From this
point they went on to develop the more general notion of a tau
function.

More recently, in a seemingly unrelated development, in studying
the geometry of vacua of
two dimensional $N=2$ supersymmetric quantum field
theories we encountered a set of equations \topatop\ which captures
Berry's curvature for the vacuum subsector of the theory.  These
equations
which were derived using
the $N=2$ algebra and relied heavily on the topological
nature of these theories were called {\it
topological-anti-topological}
or $tt^*$ equations.  Some aspects of these computations were
recently
reinterpreted \newindex\ as a way to define and compute a new
supersymmetric
index: $Tr(-1)^F F exp(-\beta H)$.  In the simplest example,
with just two vacua, the $tt^*$ equations turn out to be equivalent
to Painleve III.

This appearance of the same equation in two seemingly
unrelated physical problems raises the question of whether
there is any deeper relation between the two.
Surprisingly it turns out that, with
a suitable choice of coordinates, the
$tt^*$ equations for an arbitrary $N=2$ theory with $n$ vacua
are identical to the equations written
in \japanese\ for the $n$ point spin correlation function for massive
Ising model!
It is the aim of this paper to shed some light on this connection.
In
particular the tau function which is naturally defined in the Ising
model, should find an interesting interpretation in the $N=2$
context.
We find that the $N=2$ analog of the tau function
is another `index'.

In section 2 we review the work on the Ising model following
\japanese .
In particular we see how to rephrase the spin correlation
computation as solving a certain cohomological
problem.  This cohomological description is very useful in connecting
it to $N=2$ supersymmetric theories as the vacua in these
theories are also characterized cohomologically.  In section 3
we review the $tt^*$ equations, and also present an alternative
(less rigorous) derivation for  them.  We then write these
equations in coordinates suitable for making contact with those
of the Ising model.
In section 4 we establish a more direct link between the Ising model
and the $N=2$ theories, by considering the supersymmetric quantum
mechanical
Landau-Ginzburg models.  This provides a dictionary between the
two theories and
makes manifest why apriori the $tt^*$ equations
are the same as the equations appearing for Ising model. Once this is
established we use the Ising model correlators
to compute the exact
ground state wave function for certain SQM problems.
In section 5 we discuss  the differential geometrical aspects
of the tau function which can be constructed from any solution of the
$tt^*$
equations.  In section 6
we  discuss how to obtain the tau function via path integrals
 which paves the way to interpreting tau-function as a new
supersymmetric
index.  This is done by defining it as an integral of the
toroidal partition function (with insertion of $F^2$) integrated
over the moduli of torus.
 In this way we are led to a generalization of Ray-Singer torsion
in the context of {\it loop space} of Kahler manifolds. This also
turns out to be related to computations of moduli dependence
of gravitational and guage coupling constants for string theory
\oneloop .

Finally in section 7 we present our conclusions.

 \newsec{Ising Correlations as a Problem in Hodge Theory}

The theory of N=2 supersymmetric
systems is closely related to that of the $2d$
Ising  model \ising. This is particularly evident from the
`monodromic'
approach advocated by the Japanese school \japanese.
Solving the Ising model is equivalent to computing some
 cohomology which  is a canonical
model for the $Q$--cohomology of a `generic'
 N=2 model. To establish the
dictionary between the two theories, we briefly review the main
results of ref.\japanese\ from a viewpoint appropriate to making
contact with N=2 susy. (Experts may wish to skip
this section). This will also make clear the relation of the Ising
model to Hodge theory \hodge. A more physical link between the two
theories will be discussed in sect. 4.

\subsec{The Ising Model}

The starting point is the well known fact that,
in the scaling limit, the Ising model
 is just a free massive Majorana field $\Psi(z)$ satisfying
the Euclidean Dirac equation
\eqn\dirac{(\dsl - m)\Psi=0}
where
$$\dsl=\left(\matrix{0 & \partial\cr \overline{\partial} &
0\cr}\right),\qquad \Psi=\left(\matrix{\psi_+\cr \psi_-\cr}\right).$$
If $\Psi$ is a solution to \dirac ,
 $\Psi^\star\equiv-\sigma_1\Psi^\ast$ is also a solution.
Therefore the space of solutions
to \dirac\ admits a {\it real structure}:
A solution is said to be {\it real} if $\Psi=\Psi^\star$. Of course,
this
is just the (Euclidean) Majorana condition.  From this point
on we will usually set, by a choice of units, $m=1$.

In addition to $\Psi(z)$, one considers two other fields
\disorder: The order operator
$\mu(z)$ and the disorder one $\sigma(z)$. Insertion of these
operators
creates a cut in the plane, around which the fermi field $\Psi(z)$
picks
up a sign as it is transported around it.
 These fields can be written
(non--locally) in terms of the
fundamental fermion $\Psi$ \japanese . One has
$\mu=\exp[\varrho]$ with $\varrho$ a (non--local) bilinear in $\Psi$.
Instead $\sigma$ has the form $\psi \exp[\varrho]$
with $\psi$ an operator linear
in $\Psi$. Then $\mu$ is an even element of the Grassman algebra
generated by $\Psi$ while  $\sigma$ is
odd. By definition they satisfy
$\langle\mu\rangle=1$ and
$\langle\sigma\rangle=0$. The precise form of these operators
can be extracted from the path--integral
representation of the order/disorder correlation functions.
 These functions are equal \refs{
\pathint\marino} to the functional integral
over a free fermion $\Psi$ having a prescribed
monodromy around the points $z_i$
where the order/disorder operators are inserted.
After doubling the model by making
$\Psi$ complex, this boundary condition
can be implemented by coupling its Fermi
current to a flat Abelian connection
$A_\mu$ whose  holonomy equals the given
monodromy. In this way one gets \eg\
\eqn\disorder{\mu(z)=
N\exp\left[i\int d^2w \overline\Psi(w) \Vsl{A}(w;z) \Psi(w)\right],
}
where $N$ is some normalization coefficient and
$A_\mu(w;z)$ is a flat connection
on the punctured plane with
$$\oint_CA=\pi \qquad {\rm
for\ all\ paths}\ C\ {\rm enclosing}\ z.$$
The disorder operator $\sigma(z)$ is defined
similarly, except that one couples $A_\mu$
to the axial current \pathint\ (the
procedure is rather subtle since the
axial current is not conserved). After the
path integral is twisted in this way one
(and only one) normalizable zero--mode for
$\Psi$ appears\foot{The zero--mode wave
functions are discussed in detail below.}.
Since the zero--mode absorbs a fermion,
$\sigma(z)$ becomes an {\it odd}
element of the field algebra.

The relation of the order/disorder fields to the fundamental fermion
is
conveniently summarized by their OPE with $\Psi(z)$.
The operator expansions can be
extracted from the above functional representation \pathint .
Consider
the product $\Psi(z)\sigma(w)$ as $z\rightarrow w$. By definition of
disorder
operator \disorder, this product changes sign as we transport $z$
around $w$.
 Moreover, as a
function of $z$, it satisfies the Dirac equation \dirac. Let
$\phi_i(z)$
be a basis of solutions
with the right monodromy around
the origin. Then we can expand the above product in this basis.
In particular we can use use a basis with definite angular
momentum\foot{In
the first quantized sense.}. For each $l\in {\bf Z}$
there are two such functions $\varphi_l(z)$ and
$\varphi^\star_{-l}(z)$,
whose explicit expression in terms of modified Bessel functions reads
\eqn\bessel{\varphi_l(z)\equiv
\left(\matrix{\zeta_l(z)\cr
\zeta_{l+1}(z)\cr}\right)=
\left(\matrix{\exp\left[i(l-\half)\theta\right]\, {\rm
I}_{l-\half}(2r)\cr \exp\left[i(l+\half)\theta\right]\, {\rm
I}_{l+\half}(2r)\cr}\right)  }
where $z=r\exp[i\theta]$.  Note that for small $z$
$$\zeta_l \sim z^{l-{1\over 2}}.$$

Then, the basic OPE are \japanese\
\eqn\OPE{\eqalign{\Psi(z)\sigma(w) &=\textstyle{i\over 2}\mu(w)
\big(\varphi_0(z-w)+\varphi^\star_0(z-w)\big)\cr
&+\half\sum_{l=1}^\infty
\big[\mu_l(w)\varphi_l(z-w)+\mu_{-l}(w)\varphi_l^\star(z-w)\big]\cr
 \Psi(z)\mu(w) &=\half\sigma(w)
\big(\varphi_0(z-w)-\varphi^\star_0(w-z)\big)\cr
&-\half\sum_{l=1}^\infty
\big[\sigma_l(w)\varphi_l(z-w)-\sigma_{-l}(w)\varphi_l^\star(z-w)\big
].\cr} }
Here $\sigma_l(z)$ and $\mu_l(z)$ are `descendants' of the basic
order/disorder fields. In particular,
$$\mu_1= i\partial \mu,\quad \mu_{-1}=- i\bar\partial\mu,\quad
\sigma_1= \partial\sigma,\quad \sigma_{-1}
= \bar\partial\sigma,$$
as is easily seen by rewriting \OPE\ in terms of free fields.

In the space of solutions we have a natural inner product,
  corresponding to the norm
in the (first quantized) Hilbert space (where we take,
for later convenience, a factor
of $1/2 \pi$ as part of the definition of the integral)
\eqn\norm{\Vert\Psi\Vert^2=\int d^2z \big(|\psi_+|^2+|\psi_-|^2\big).
}

The main problem in the theory is to compute the wave function
 induced by the
insertion of one disorder and many order operators \japanese\
\eqn\wavefunction{\Psi^{(j)}(z)=
{\langle \Psi(z)\mu(w_1)\dots\mu(w_{j-1}) \sigma(w_j)
\mu(w_{j+1})\dots\mu(w_n)\rangle\over
\langle\mu(w_1)\dots\mu(w_n)\rangle}.  }
Once these functions are known, the order/disorder
correlations can be computed using  \OPE . In fact, as $z\rightarrow
w_j$
we have
\eqn\diaglim{\eqalign{\Psi^{(j)}(z)&={i\over
2}\big[\varphi_0(z-w_j)-\varphi_0^\star(z-w_j)\big] - {i\over 2}
\varphi_1(z-w_j)\, \partial_{w_j}
\log\langle\mu(w_1)\dots\mu(w_n)\rangle \cr
&+ {i\over 2} \varphi_1^\star(z-w_j)\,
\partial_{\overline{w}_j}\log\langle\mu(w_1)\dots\mu(w_n)\rangle
+O(|z-w_j|^{3/2})\cr} }
so the function
\eqn\taufunct{\tau(w_1,\dots, w_n)\equiv
\langle\mu(w_1)\dots\mu(w_n)\rangle,}
can be recovered by quadrature from $\Psi^{(j)}$.  From the path
integral point
of view, $\tau$ is just the determinant
$$\tau ={\rm Det}[\Dsl-m],$$
where $\Dsl$ is the Dirac operator coupled to the
connection $A^{(n)}_\mu$
induced by the insertion of the order fields.

On the other
hand, as $z\rightarrow w_k$ ($k\not=j$) we get
\eqn\knotj{\Psi^{(j)}(z)={1\over
2}\big[\varphi_0(z-w_k)-\varphi_0^\star(z-w_k)\big] \tau_{jk}+
{\rm non\ singular},}
where
$$\tau_{jk}\equiv
{\langle \mu(w_1)\dots \sigma(w_j)\dots
\sigma(w_k)\dots\mu(w_n)\rangle
\over \langle\mu(w_1)\dots\mu(w_n)\rangle}.$$
Since $\sigma$ is a Grassman odd operator
$\tau_{jk}$ is a real skew--symmetric matrix.
This is obvious since in view of \OPE\ the correlation
function of two $\sigma$'s in the presence of $\mu$'s can
be obtained by taking the leading singularity of the 2--point
function
$\langle\Psi(w_j)\Psi(w_k)\rangle_{\rm A}$, where
$\langle\cdots\rangle_{\rm A}$ denotes expectation values
with the fermions coupled to $A_\mu^{(n)}$.  Similarly
$\langle\sigma(w_1)\dots\sigma(w_n)\rangle$ can be computed by
applying Wick's theorem. This gives the
identity \japanese\
$$\langle\sigma(w_1)\dots\sigma(w_n)\rangle=
\tau\, {\rm Pf}[\tau_{ij}].$$

To get the wave functions $\Psi^{(j)}(z)$ one looks for the
multivalued solutions to \dirac\ having
singularities at the branching points
$w_k$ as predicted by the OPE, and
behaving as $O\left(e^{-2|mz|}\right)$ at infinity.
To find such functions
is the `cohomological' problem which we will elaborate
in the next subsection.

There is another Ising quantity which plays a crucial
 role in the susy case. Let
$L$ be the (first quantized) angular momentum. It is a first order
differential operator acting on $\Psi$.
 The wave functions $\Psi^{(j)}$ have no
definite angular momentum since the
 insertion of order/disorder operators breaks
rotation invariance. Nevertheless,
$L\Psi^{(j)}$ is again a solution of
\dirac\ with the same monodromy properties
as $\Psi^{(j)}$ and slightly more
singular at the $w_k$, as is easily
seen by acting with this differential
operator on the local expansions \diaglim\knotj.
We define the matrix\foot{Notice that this
matrix is independent of the choice
of the origin in the $z$--plane.}
\eqn\qmatr{Q_{ j i}=-\langle\Psi^{(j)}|L|\Psi^{(i)}\rangle.  }
This $Q$--matrix will correspond to
an `universal' supersymmetric index in the $N=2$ context.

\subsec{The Cohomological Problem}

In order to emphasize the geometrical
meaning\foot{Our discussion here can
be seen as the {\it massive} analog
of the well--known fact that, on a
Ricci--flat K\"ahler manifold, the
K\"ahler--Dolbeault equation is equivalent
to the massless Dirac equation
\hitchen\ Th.2.2.
(See also sect.3 of \WKD ). In
our case the relevant Calabi--Yau space is just the flat complex
plane.
It would be interesting to extend the
massive result to higher dimensional CY
spaces.}
of the Euclidean Dirac equation, let us replace $\Psi(z)$ by
the 1--form
\eqn\waform{\varpi=\psi_+(z)dz+\psi_-(z)d\bar z.  }
The map $\Psi\mapsto\varpi$ is very natural. First of all, it
is consistent with the Hilbert space structure \norm
\eqn\hilbspace{\Vert\Psi\Vert^2=\int \varpi\wedge\ast\varpi^\ast}
(again a factor of $1/2\pi$ is taken as part of the
definition of the integral).
Moreover, in terms of $\varpi$, the
real structure becomes the standard one:
$\Psi$ is real in the Majorana sense
if $ i\varpi$ is real in the usual sense.
Finally, the
form $\varpi^{(j)}$ associated to
 the basic wave--function \wavefunction\ is
regular everywhere. More precisely, it is regular when pulled back on
the
branched cover on which $\Psi$ is univalued.  This branched cover
is just a hyperelliptic surface which is branched over the points
where the spin operators are inserted.
In fact, $dz/\sqrt{z-w_k}$ is regular at $w_k$, as is seen by writing
it
in terms of the local uniformizing parameter $\varsigma^2=(z-w_k)$
of the hyperelliptic surface.
This regularity condition encodes the essential part of the OPE \OPE.
 Let  $\chi$ be the
monodromy defined by  picking up a minus sign when going around
the branch points $w_k$.
Then $\varpi^{(j)}$ may also be seen as a form taking
value in the flat vector bundle $V_\chi$ defined by the monodromy
$\chi$.

Consider the operators
$\bdi$ and $\CD$ acting on forms as
\eqn\defcha{\bdi\alpha=\bar\partial\alpha+\,
dz\wedge\alpha\qquad  \CD\alpha=
\partial\alpha+\, d\bar z\wedge\alpha.  }
They satisfy
$$\bdi^2=\CD^2=\bdi \CD+\CD \bdi=0$$
so $\bdi$ (resp. $\CD$) is a cohomology operator.
They are consistent with the
inner product in the sense that,
 for $\bdi$--closed {\it one} forms $\alpha$, and $\beta$, the
product
$\int\alpha\wedge\ast\beta$ depends only on their classes.

In terms of forms, Eq.\dirac\ reads
$$\bdi\varpi=\CD\varpi=0.$$

Let $\Lambda$ be the linear operator which
vanishes when acting on 0 or 1 forms
and maps the 2--form $i a(z) dz\wedge d\bar z$
to the zero--form $a(z)$.
Then one has the K\"ahler identity (familiar from
the study of K\"ahler manifold), which can be easily
verified
$$\bdi^\dagger= i [\Lambda, \CD].$$
 Using this identity, we
rewrite the Dirac equation as
\eqn\KDE{\bdi^\dagger \varpi=\bdi\varpi =0. }
The first equation is equivalent to the first
component of the Dirac equation and the second equation
is equivalent to the second component.
Hence the (normalizable) solutions to \dirac\ are just the
`harmonic' representatives of the $\bdi$--cohomology in a space
of forms satisfying suitable boundary conditions.

By standard Hodge theory, these
solutions are in one--to--one correspondence
with the $\bdi$--cohomology classes in the same space.
Complex conjugation interchanges
$\bdi$ with $\CD$ and thus maps the
$\bdi$--cohomology into the $\CD$--cohomology (equivalently the
$\bdi^\dagger$ one). Notice that a `harmonic'
 form $\varpi$ can be seen either as a
 $\CD$-- or a $\bdi$--cohomology class,
 the two classes being related by the reality
structure.

In particular  $\varpi^{(j)}$ is a
 `harmonic' 1--form representing a cohomology
class in the space\foot{A priori one adds also the condition of
fast decay at infinity. It can be shown (\compact\ App.B)
that such
restriction is immaterial in cohomology,
\ie\ each cohomology class has a fast decaying
representative.} of  (regular) forms with coefficients in $V_\chi$.
Let
$\Omega_k$ be a basis of such `harmonic' forms. Then
\eqn\changebasis{\varpi^{(j)}=
\sum_k A^{(j)}_k \Omega_k, }
for some coefficients $A^{(j)}_k$. To get these
coefficients, it is enough to compare the
two sides of this equation in {\it
cohomology}.

{}From \waform\ we see that for a `harmonic'
one form\foot{In fact, in the present context, only one forms
can be harmonic.}
$$\varpi=  \bdi\psi_+,$$
where $\psi_+$ is the first component of the corresponding
Dirac spinor.
Since in cohomology we are free to add to $\varpi$ any
$\bdi\xi$ with $\xi$ regular, the $\bdi$--cohomology class
of $\varpi$ is completely encoded in the
{\it singularities} of $\psi_+$. The only allowed
singularities of $\psi_+$ are of the
form $a_k/\sqrt{(z-w_k)}$. So, the
$\bdi$--class of $\varpi$ is specified
by the numbers $(a_1,a_2,\dots, a_n).$
The {\it canonical} basis $\Omega_k$ is defined
by $a_i(\Omega_k)=\delta_{ik}$. In this
basis we get
\eqn\resmetric{\eta_{kj}=\int\Omega_k\wedge\ast\Omega_j= \delta_{jk},
}
This formula can be proven using residue techniques \residueapp .
Alternatively, in the present case we can give an explicit
proof of this which will also be useful for some other
applications later in this section. Consider functions $\psi_a$
and $\psi_b$ decaying exponentially fast at infinity
which, except possibly at $w_i$, are regular and around
$w_i$ pick up a sign.  Then we will show
that
\eqn\interest{\int \bdi \psi_a \wedge \ast\bdi \psi_b =\sum_{w_i}
\oint_{w_i} dz\  \psi_a \psi_b}
where $\oint_{w_i} $ is a contour integral taken in an infinitesimal
neighborhood of $w_i$ (with a factor $(1/(2\pi i)$ included).  The
proof is
quite simple.  We simply
use the definition \defcha\ and expand the above, and use the
definition of $\ast $ to obtain
$$\int \bdi \psi_a \wedge \ast\bdi \psi_b =-i\ \int  \bar{\partial}
(\psi_a
\psi_b) dz$$
Using Stoke's theorem, applied to the complex plane with
infinitesimal discs cut out around $w_i$ where $\psi$'s are singular,
we end up with equation \interest .  Now we can apply this result
to $\psi_a=\psi_{k+}$ and $\psi_b=\psi_{j+}$ we obtain the result
\resmetric .

Now let us consider the natural hermitian metric of the Hilbert space
in the
canonical basis
$$g_{i\bar j}= \int\Omega_i\wedge\ast\Omega_j^\ast.$$
Comparing with \resmetric , we see that $\Omega_j^\ast=
\sum_i g_{i\bar j}\Omega_i$, \ie\
the matrix $g$ represents the real
structure in the canonical basis. Then we must have (by conjugating
twice)
$$g^{-1}=g^\ast = g^T.$$
The Ising correlations can be extracted
 from the metric $g$ as we will now show. We can read
the cohomology classes of $\varpi^{(j)}$
from \diaglim\ and
\knotj . Comparing with \changebasis, this gives
$$A^{(j)}_k =i\half \delta_{jk}+\half
\tau_{jk}$$
Using reality of $i \varpi^{(j)}$
we obtain
$$A^{(j)*}_k=-A^{(j)}_k g_{k\bar l}$$
which leads (using reality of $\tau_{ij}$) to

$$ A^{(j)}_k g_{k\bar l}=
i\half\delta_{jl}-\half \tau_{jl},$$
which gives $A^{(j)}_k$ and $\tau_{jk}$ in terms of $g$
\eqn\ansfa{A^{(j)}_k=i (1+g)^{-1}_{jk}}
\eqn\whattauma{\tau_{jk}=(A-Ag)_{jk}=i [(1+g)^{-1}(1-g)]_{jk}. }
To complete the computation of the correlators,
we need to compute the $\tau$
function \taufunct .
Let $\Omega_k=i\bdi\psi_{k+}$. Near $w_j$,
$\psi_{k+}$ has an expansion of the
form
\eqn\expansion{
\eqalign{
\psi_{k+}(z)&
\simeq \delta_{kj}\zeta_0(z-w_j)
+\tilde Z_{kj}\zeta_1(z-w_j)
\cr &-\sum_{l}g_{k\bar
l}[\delta_{lj}\zeta_1(z-w_j)^\ast +\tilde
Z_{lj}^\ast\zeta_2(z-w_j)^\ast]
+\dots \cr}  }
where we have used the Majorana condition and the fact
that complex conjugation of the basis is equivalent to
multiplication by $g$.
Comparing with \diaglim\ we see that
($\partial_j\equiv \partial_{w_j}$)
\eqn\partau{({A\tilde Z})_{jj}=[i(1+g)^{-1}\tilde
Z]_{jj}=-\textstyle{i\over
2}\partial_j\log\tau. }
Since, as we are going to show,
the matrix $\tilde Z$ can also be expressed
in terms of $g$, the
$\tau$ function can be recovered from $g$ by quadratures
once we know how to compute $g$.

It remains to compute $g$ and
$\tilde Z$ as a function of $w_1,\dots, w_n$.
The idea is to study the differential geometry of the
hermitian metric $g$.
In this way we are led to a rather standard problem
\foot{One can also reduce to
a problem of `isomonodromic deformations' as in \japanese .
The two mathematical theories are
closely related \deligne. However, in the
present language the underlying
 ring structure is more transparent. In the susy
case the ring is the central object of interest.} in `variation of
Hodge
structures' \hodge .
There is a natural {\it metric connection} for $g$
$$D_i=\partial_{i}-A_i,\qquad
\bar D_j=\partial_{\bar j}-\bar A_j,$$
where the connection is defined by
\eqn\defconnection{\eqalign{& (A_i)_{kl}=\int
\partial_{i}\Omega_k\wedge\ast\Omega_l\cr & (\bar A_i)_{kl}=\int
\partial_{\bar i}\Omega_k\wedge\ast\Omega_l .\cr}  }
It is easy to check with this definition that $D_ig=\bar D_i g=0$.
Also, in the basis we have chosen we will now show that the
anti-holomorphic
components of the connection vanishes,
\eqn\holim{(\bar A_i)_{kl}=\int
\partial_{\bar i}\Omega_k\wedge\ast\Omega_l\equiv 0}
To prove this we apply the result \interest\ with the substitutions
$\psi_a= \partial_{\bar i}\psi_{k+}$ and $\psi_b= \psi_{l+}$.
Since the singularity of $\psi_a $ near $w_k$ is of the form
$\delta_{ik}/\sqrt{\bar{z}-\bar{w_k}}$ and that of $\psi_b$ is of the
form
$\delta_{kl}/\sqrt{ {z}- {w_k}}$, using \interest\ we end up with
$$(\bar A_i)_{kl}=\delta_{ik}\delta_{kl}\oint_{w_k} {dz\over
|z-w_k|}=0$$
Using this result and using the fact that $D_ig=0$ we get,
\eqn\canconn{A_i=-g\partial_i g^{-1}. }
{}From \expansion\ we have
$$ \partial_i\psi_{k+}=
-\delta_{kj}\delta_{ji}\zeta_{-1}(z)-
\tilde Z_{kj}\delta_{ji}\zeta_0(z)+\,
{\rm regular\ as\ }z\rightarrow w_j.$$
Using this, and evaluating the integral
\defconnection\  using \interest\ we get
$$(A_i)_{kl}=\tilde Z_{kl}(\delta_{ik}-\delta_{il}).$$
It is convenient to introduce the matrices
 $(C_i)_l^k\equiv \delta_{il}\delta_l^k$.
Then the connection is rewritten as
\eqn\otherconn{A_i=[C_i,\tilde Z].  }
Comparing with \canconn ,
 we get a formula for off--diagonal part of the matrix
$\tilde Z$ in terms of the metric $g$.
In particular, $\tilde Z^T=\tilde Z$.
Moreover we have the identity
\eqn\secondtt{[C_j,A_i]=[C_i,A_j].  }
which follows from \otherconn, the Jacobi identity, and
the fact that $C_i$ commute with one another.

We still need to compute the diagonal part of $\tilde Z$.
It is convenient to define the matrix
$\bar C_{\bar i}\equiv gC_i^\dagger
g^{-1}$. Then one has the identity
\eqn\coharg{\partial_{\bar i}\Omega_k
+(\bar C_{\bar i})_k^{\ l}\partial_{\bar
z}\Omega_l=0.  }
To show this, note that the left hand side
of the above equation is clearly `harmonic' (as
$\partial_{\bar z}$ and $\partial_{\bar i}$ commute
with $\bdi$ and $\CD$).  Moreover using the expansion \expansion , it
is
easy to see that the
the left hand side is a regular `harmonic' form (the
singularities cancel between the two terms).
Since it is regular (i.e. it has no singularities),
it represents the trivial class,
and hence vanishes identically.
Setting to zero the coefficient of $\zeta_1$, we get
\eqn\delZ{ \bar\partial_{i}\tilde Z
= -\bar C_{\bar i} +C_i^\dagger. }
Let us compute the curvature $R_{i\bar j}$ of the above connection.
 Recalling that
$\bar A=0$ one has
$$R_{i\bar j}=\partial_{\bar j}A_i= [C_i, \partial_{\bar j}\tilde
Z]=- [C_i,\bar C_{\bar j}].$$
Using the explicit form of the connection \canconn ,
 this becomes a differential
equation for the metric $g$
\eqn\diffeqns{\partial_{\bar j}
(g\partial_i g^{-1})= [C_i,\bar C_j].  }
This is the basic equation governing the correlation functions
(if we restored $m$, there will appear a factor $m^2$ on the rhs
of the above equation).

Let us introduce some short--hand notation,
\eqn\shorthand{C=\sum_k w_k\, C_k,
\quad \Gamma= \sum_k C_k\, dw_k, \quad \bar C=
g C^\dagger g^{-1},\quad \bar\Gamma= g \Gamma^\dagger g^{-1}.  }
Then consider the expression
$$(L+ C\partial_z-\bar C\partial_{\bar z})_k^{\ l}\Omega_l,$$
where $L=w_i\partial_i -{\bar w_i}\partial_{\bar i}+\half\sigma_3$
 is the angular
momentum operator.  Then again this is a  `harmonic' form which
is regular in the sense of the discussion after eq.\hilbspace\ and
hence it is of the form $Q_k^{\ l}\Omega_l$
for some matrix $Q_k^{\ l}$.  Comparing with $\qmatr$, we see that
this matrix is the one defined in $\qmatr$
(note that with the flat metric \resmetric\ we can lower or raise
indices of $Q$ and that $Q$ is
antisymmetric).  Moreover, using \defconnection\ and \holim\ we get
\eqn\isingQ{Q=-\sum_k w_k A_k=
[\tilde Z, C]\equiv \sum_k w_k \, g\partial_k g^{-1}. }
{}From \secondtt\ one has $[C,g\partial g^{-1}]=[\Gamma, Q]$.
 This allows to write
the connection in terms of $Q$ alone,
$$(g\partial g^{-1})_{kl}= Q_{kl}{d(w_k-w_l)\over w_k-w_l}.$$
This fact can be used to reduce the number
 of independent equations in \diffeqns .

{}From \partau\ we have
$$ \partial\log\tau= -2\tr[\Gamma (1+g)^{-1}\tilde Z]$$
now we use the fact that
$$\tr M = {1\over 2}(\tr M +\tr M^t)$$
for any $M$ and
use the fact that $\tilde Z^t=\tilde Z$, $\Gamma ^t=\Gamma$
and that $g^t=g^{-1}$ and \otherconn\ to get
$$\partial\log\tau = \ \tr  (\Gamma (1+g)^{-1}\tilde Z +\tilde Z
(1+g^{-1})^{-1}
\Gamma )$$
$$= \ \tr (\Gamma [(1+g)^{-1}+(1+g^{-1})^{-1}]\tilde Z +
(1+g^{-1})^{-1}[\Gamma
,\tilde Z]) $$
$$= \ \tr \Gamma \tilde Z +\tr (1+g^{-1})^{-1}(-g\partial g^{-1})$$
$$=  \ \tr \Gamma \tilde Z + \tr (1+g)^{-1}\partial g$$
\eqn\gan{= \ \tr \Gamma \tilde Z + \partial \tr \log (1+g)}

We will now show that
\eqn\neas{\tilde Z= - \bar C + C^\dagger+
\half [Q,\tilde Z]+\, {\rm off-diagonal}.}
Let $\Delta $ be the difference between lhs and the rhs of the above
equation.
Now let us compute $\bar \partial_i \Delta$.  We will need
\eqn\curq{\bar\partial_i Q = [C, \bar C_i] }
which follows from \diffeqns\ and
\eqn\curqq{\bar \partial_i \bar C =\bar C_i -[Q,\bar C_i]}
To show \curqq\ we first use \secondtt\ to get
$$[Q,C_i]=-[C,g\partial_i g^{-1}].$$
We then note that $Q^*=-g^{-1}Qg$  which follows from rotational
invariance of the theory  $Lg=0$ and $g^*=g^{-1}$.  Taking the
complex conjugate of the above equation and noting that
$C_i=C_i^*=C_i^\dagger$
we get
$$g^{-1}[Q,\bar C_i]g=-[g^{-1}\bar \partial_i
g,C^\dagger]=-g^{-1}(\bar
\partial_i \bar C -\bar C_i)g$$
which gives equation \curqq .

Finally we are in a position to compute $\bar \partial_i \Delta$ .
We have
from \delZ\
$$\bar \partial_i \Delta =-(\bar C_i- C_i^\dagger)+ \bar
\partial_i\bar C
- \bar\partial_i C^\dagger -{1\over 2}[\bar \partial_i Q, \tilde Z]+
{1\over 2} [Q,\bar C_i]-{1\over 2} [Q,C_i^\dagger]$$
Now using \curq\ and \curqq\ (and noting $\bar \partial_i
C^\dagger =C_i^\dagger$) we have
$$\bar \partial_i \Delta =- {1\over 2}[[C,\bar C_i],\tilde Z]-
{1\over 2} [Q,\bar C_i]-{1\over 2 }[Q,C_i^\dagger]$$
Using Jacobi identity on the first term and using \isingQ\ we get
$$\bar \partial_i \Delta ={1\over 2}[[\bar C_i,\tilde Z],C]-{1\over
2}
[Q,C_i^\dagger]$$
The rhs of the above equation has vanishing diagonal elements, since
$C$
and $C_i^\dagger$ are diagonal matrices and $[M,D]$ has vanishing
diagonal
elements for any matrices $D$ and $M$ with $D$ diagonal.  So we have
now shown that the difference between
the diagonal elements of both sides of \neas\
is at most a holomorphic function in $w_i$'s.  Since everything falls
off exponentially fast as any of $w_i\rightarrow \infty$, the
diagonal
elements of $\Delta$ die off at least exponentially fast in this
limit,
and this implies, since diagonal elements
of $\Delta$ are holomorphic, that they are identically zero.  This
concludes
showing \neas .

Now let us define $\tilde \tau$ by
$$\log \tilde \tau= \log \tau - \log {\rm det}(1+g),$$
then using \neas\ and \gan\ and noting that $\Gamma$ is a diagonal
matrix (and using the reality of $\log \tau$) we get
\eqn\taudiff{d\log\tilde\tau=
-  \tr[\bar C\Gamma]+ \tr[C^\dagger\Gamma]
+\half\tr\big(Q g\partial g^{-1}\big)+\, {\rm c.c.}     }
We stress that the rhs of \taudiff\
is a closed form for all solutions to
\diffeqns . So to each solution we
can associate (locally) a $\tau$--function.
Of course, only a particular solution of
 these equations corresponds to the
actual correlation functions for the Ising model.

\newsec{The Geometry of N=2 Ground States}

\subsec{The \ttstar\ Equations}

In this section we review the main results of ref.\topatop .
We consider a general N=2 susy model in two dimensions.
A remarkable fact about such models is that they can be `twisted'
into a
topological field theory (TFT)
 \TFT . This is done by gauging the Fermi
number $F$ with the gauge field set equal to one--half the
spin--connection.
After the twisting, two supercharges $Q^+$
and $\bar Q^+$ become scalars and are
consistently interpreted as BRST charges. The observables $\phi_i$
of the TFT are the operators commuting with these supercharges modulo
those that can be written as $Q^+$ or $\bar Q^+$--commutators. In
technical terms, the $\phi_i$'s correspond to the $Q^+$--cohomology
in the space of quantum operators.
In the same way the physical states are
associated to the $Q^+$--cohomology in the Hilbert space, \ie\
$$Q^+|{\rm phy}\rangle=\bar Q^+|{\rm phy}\rangle=0,
\qquad |{\rm phy}\rangle \sim |{\rm phy}\rangle+Q^+|\xi\rangle+
\bar Q^+|\eta\rangle.$$
The topological observables $\phi_i$ generate a commutative\foot{If
some
$\phi_i$ have odd $F$, $\CR$ is only {\it graded--commutative}. To
keep
notations as simple as possible,
we assume all $\phi_i$ to have even $F$. The
extension to the general case is straightforward.} ring $\CR$
\eqn\chiring{\phi_i\phi_j =C_{ij}^k\phi_k.  }
The ring $\CR$ is called the
{\it chiral ring} \chiralring\ since, in the original N=2 model, the
fields $\phi_i$ commuting with $Q^+$ and $\bar Q^+$ are just the
chiral ones.
The two-point function on the sphere serves
as a metric for the topological theory
$$\eta_{ij}=\langle\phi_i\phi_j\rangle.$$
If we take the space to be a circle,
the physical states are in one--to--one correspondence with the
physical
operators. Explicitly, the relation reads
$$|\phi_i\rangle=\phi_i|1\rangle,$$
where $|1\rangle$ is a canonical state whose
wave--functional is obtained by filling the circle with a disk
and performing the topological path integral over this disk\foot{For
a
detailed  discussion, see \eg\ \topatop .}. Then the state
$|\phi_i\rangle$ is represented by the path integral  with the
 topological observable $\phi_i$
inserted at some point on the disk. This shows
that
\eqn\ringonstates{\phi_i|\phi_j\rangle= C_{ij}^k|\phi_k\rangle.  }
It turns out that
all the correlation functions
can be computed in terms of $C_{jk}^l$ and
$\eta_{jk}$ \TFT .

What is the connection of
the topological model with the original untwisted N=2
model? Well, if we put our N=2 model
 on a  (say) flat periodic cylinder, the
spin--connection vanishes and hence the twisting does not change the
functional measure.
In this case, the only difference between the two theories is in the
definition of the observable and physical states.
In the twisted case they are
required to be $Q^+$--closed. But then, if we restrict ourselves to
$Q^+$--closed objects in the $N=2$ theory the results for the twisted
and untwisted models are equal.

Now consider the ground states $|i\rangle$ of the untwisted theory.
They satisfy,
\eqn\vacuaeq{ Q^+|i\rangle=\bar Q^+|i\rangle
=\big(Q^+\big)^\dagger|i\rangle=
\big(\bar Q^+\big)^\dagger|i\rangle=0.  }
Comparing with the topological theory, we see
the vacua are nothing else than the
`harmonic' representatives of the BRST--classes.
Hence there is a one--to--one
correspondence between the N=2 vacua and the TFT
physical states $|\phi_i\rangle\mapsto|i\rangle$.
For future reference, we describe the
path integral realization of this map. Just
glue to the boundary of the disk of perimeter
$\beta$ a flat cylinder with the same perimeter and of length $T$.
This
does not change the topological state;
but from the N=2 viewpoint
this is the state $\exp[-HT]|\phi_i\rangle$.
As $T\rightarrow\infty$,
$\exp[-HT]$ becomes a projector on
the ground states, and we get the unique
vacuum in the $|\phi_i\rangle$ topological class.
 In particular from \ringonstates\ we have
$$\phi_i|j\rangle= C_{ij}^k|k\rangle\quad
{\rm mod. \ positive\ energy \
states}.$$

Of course, we could as well have gauged $F$ with
{\it minus} one--half the spin connection. In
this case the other two supercharges
$Q^-\equiv (Q^+)^\dagger$ and $\bar Q^-\equiv (\bar Q^+)^\dagger$
are interpreted
as BRST charges.
This is the so--called {\it anti--topological} model \topatop .
The new observable operators $\bar\phi_i$
are the CPT conjugates (in the N=2
sense) of the $\phi_i$'s. Then,
$$\bar\phi_i \bar\phi_j =C_{ij}^{\ast\, k}\bar\phi_k.$$
Again there is a natural isomorphism of $\overline{\CR}$ with the
anti--topological physical states\foot{$|\bar 1\rangle$ is defined in
analogy with $|1\rangle$ but this time with respect to the {\it
anti--topological} path integral.}
$|\bar\phi_i\rangle\equiv\bar\phi_i|\bar 1\rangle$.

{}From \vacuaeq\ we see that
the ground states are `harmonic' representatives of both $Q^+$ and
$Q^-$
cohomology. Then the
above construction gives us {\it two}
preferred basis in the space of ground
states:
The topological (or {\it holomorphic}) one $|i\rangle$,
and the anti--topological (or {\it anti--holomorphic}) one
$|\bar i\rangle$. The object of primary
interest is the hermitian metric\foot{We
adhere to the topological conventions.
Then the adjoint of $|\phi_i\rangle$ is $\langle\bar\phi_i|$
not $\langle\phi_i|$.}
\eqn\gmetric{g_{i\bar j}=\langle \bar j| i\rangle,  }
which intertwines between the two `topological' bases. It is an
highly
non--trivial quantity.
On the contrary,
 $\langle i|j\rangle$ is an `elementary' topological object
\eqn\etametric{\langle i|j\rangle\equiv
\langle\phi_i|\phi_j\rangle=
\langle\phi_i\phi_j\rangle_{\rm top.}=\eta_{ij}. }
Comparing \gmetric\ and \etametric ,
we see that the reality structure acts on
the vacua as the matrix $g\eta^{-1}$, then
CPT implies
\eqn\greality{g\eta^{-1}(g\eta^{-1})^\ast=1.  }

Many interesting non--perturbative
information about the N=2 model can be extracted
from $g$ (see \eg\ \refs{\topatop, \newindex, \sigmamodels,
\polymers,
\veryold, \bill}). The main point here is to get a set of
differential equations for $g$. As in sect. 2,
the idea is to study its differential
geometry. These equations (known as the \ttstar\ equations)
were proven with different techniques in
refs.\topatop, \newindex, and \veryold . Here we give a very quick
(and
somewhat unrigorous\foot{People
concerned with questions of
rigor should refer to \topatop .})  argument.

The precise framework is the following.
 We consider a family of N=2 models whose action has the
form
\eqn\peraction{S= S_0+\sum_i t_i \int
\{Q^-,[\bar Q^-,\phi_i]\}+ \sum_i \bar t_i
\int \{Q^+,[\bar Q^+,\bar\phi_i]\},  }
where $t_i$ are couplings\foot{From the TFT viewpoint it may be
natural to sum  in \peraction\ on all topological fields.
However from the QFT
viewpoint it is safer to sum only over {\it marginal} and {\it
relevant} operators which lead to renormalizable field theory.
This `conservative' assumption is implicit throughout
the paper.}  parametrizing the family,
and we look for the function $g_{i\bar j}(t,\bar t)$.

As before we introduce the {\it metric}
connection $D_i=\partial_i-A_i$
and $\bar D_i=\bar\partial_i-\bar A_i$, where
$$(A_i)_{ab}=\langle a|\partial_i|
b\rangle,\qquad (\bar A_i)_{ab}=\langle
a|\bar\partial_i|b\rangle
.$$
Clearly $g$ is covariantly constant with respect to these
connections.

To get the differential equations for $g$ we need to
compute the curvature $R_{i\bar j}$ of the connection.
Using first order perturbation theory (Eq.(43.6) of \principles ), we
have
$$D_i|k\rangle=-H^{-1}\CP\partial_i H|k\rangle\equiv H^{-1}
\CP\{Q^-,[\bar Q^-,\soint\phi_i]\}|k\rangle,$$
where $\CP$ is the projection on states
of {\it positive} energy and $\oint$ denotes
integration over the circle
in which we quantize the model.
Note that by definition $\langle l| D_i|k \rangle =0$ which implies
that
$$\bar D_j (\langle l| D_i|k \rangle)=(\bar D_j \langle l|) D_i|k
\rangle
+ \langle l|  \bar D_j(D_i|k \rangle)=0$$
which implies that
$$(\bar D_j \langle l|) D_i|k \rangle =- \langle l|  \bar D_j(D_i|k
\rangle)$$
Using this and the similar result with $D$ and $\bar D$ exchanged we
get
$$\eqalign{(R_{i\bar j})_{kl}&=\langle l|
\{Q^-,[\bar Q^-,\soint\phi_i]\}\CP H^{-2}
\{Q^+,[\bar Q^+,\soint\bar\phi_j]\}|k\rangle\cr &-
\langle l|\{Q^+,[\bar Q^+,\soint\bar\phi_j]\}\CP H^{-2}
\{Q^-,[\bar Q^-,\soint\phi_i]\}|k\rangle.\cr}$$
Consider \eg\ the first term.
Notice that only $P=0$ intermediate states contribute (there is
no momentum flowing in).
Now, using the fact that the vacua
are annihilated by $Q$'s and commuting $Q^-$ and $\bar Q^-$ across
the other
operators, and using the susy algebra (with $P=0$), we get
$\langle l|\oint \phi_i\CP \oint\bar\phi_j|k\rangle$.
The second term is
obtained by $\oint\phi_i\leftrightarrow \oint\bar\phi_j$.
Given that
$$\CP= 1-|k\rangle \eta^{kl}\langle l|,$$
the curvature becomes
\eqn\gcurvature{(R_{i\bar j})_{kl}=
\langle l|\big[\soint \phi_i,\soint\bar\phi_j\big]|k\rangle-
\beta^2[C_i,\bar C_j]_{kl},  }
where $\beta$ is the perimeter of
the circle, and the
matrices $C_i$ and $\bar C_j$ are defined by\foot{These
equations are understood to hold up
to states of positive energy.}
$$\phi_i|k\rangle= (C_i)_k^{\ l}|l\rangle, \qquad
\bar\phi_j|k\rangle= (\bar C_j)_k^{\ l}|l\rangle.$$
The first term in the rhs of \gcurvature\
vanishes since $\phi_i$ and $\bar\phi_j$ commute at
{\it equal time}.
We can now choose the holomorphic guage which happens
to be just the topological basis for the vacua.  In this case
we have $\bar A_i =0$.  Here we used the fact, manifest from
\peraction , that a variation of $\bar t$
does not change the $Q^+$--class of a state).
Then the connection has the
canonical form \canconn , $A_i= - g\partial_i g^{-1}$.
 On the other hand, comparing with
the topological (resp. anti--topological) theory
\ringonstates ,
we get
\eqn\covholo{\eqalign{&(C_i)_k^{\ l}\equiv C_{ik}^{\ l},
\qquad \bar C_j= g C^\dagger_j g^{-1}\cr
&\Rightarrow D_i\bar C_{\bar j}=\bar D_{\bar i}C_i=0.\cr}   }

Putting everything together, we get the
\ttstar\ equations for the metric $g$
\eqn\ttstareq{\bar\partial_j(g\partial_i g^{-1})
=\beta^2[C_i, gC^\dagger_j g^{-1}].  }
Notice that the Ising equations \diffeqns\
are just a special instance
of the \ttstar\ equations, corresponding
to a chiral ring $\CR$ with
$$C_{ij}^{\ k}=\delta_{ij}\delta_j^{\ k},
\qquad (\beta\leftrightarrow m).$$
Below we shall see that this special
case is in fact `generic' and
is true for arbitrary $N=2$
theory (by a choice of coordinates). The same line of argument as in
the
computation of
curvature also shows that
\eqn\otherttstar{D_jC_i=D_iC_j\quad\Rightarrow\quad
\partial_i C_j-\partial_j C_i+[g\partial_i g^{-1},C_j]-
[g\partial_j g^{-1},C_i]=0   }
which in the Ising case will be equivalent to \secondtt .
To complete the
analogy, we note the equality
\eqn\Zmatrix{\tilde Z=Z+C^\dagger,   }
where $Z$ is the matrix defined for an
arbitrary N=2 model in Appendix C of
\topatop. There it is also discussed
its connection to the {\it flat
coordinates} of TFT \flatcoordinates .

On the space of couplings there
is a special vector $v=v^i\partial_i$
which generates an infinitesimal change of scale. Then
the renormalization group (RG) is
generated by the differential operator
$$\CL_v+\overline\CL_{\bar v}+
\ D-{\rm term\ beta\ function},$$
where $\CL_v$ denotes the (covariant)
Lie derivative, acting on forms as
$$\CL_v= D\, i(v)+ i(v)\, D, \qquad
({\rm here}\ D\equiv dt^i D_i).$$
The ring $\CR$, being
topological, is not renormalized. Then the RG group should act by
ring
automorphisms, \ie\ there exists a
matrix $Q$ such that the one--form
$\Gamma$ is defined in \shorthand\ transforms as
\eqn\RGQ{\CL_v \Gamma =\Gamma+[Q,\Gamma],\qquad
 \overline\CL_{\bar v}\bar \Gamma=\bar\Gamma-[Q,\bar \Gamma].  }
The first term on the right hand side comes from the fact that
the only effect of the RG group
is to rescale the topological operators by an overall term $e^\tau$
where
$\tau$ is the RG time.  This
is a simple consequence of dimensional analysis of the (F-term part
of the)
action which remain unrenormalized.  Needless to say the first term
could be
gotten rid of by multiplying $\Gamma $ with $e^{-\tau}$.

Since the RG flow preserves the topological metric $\eta$,
$$(Q\eta)+(Q\eta)^T=0.$$
Given that $D$ is also a metric connection for $\eta$, we have
$D=D_\eta+T$, where $D_\eta$ is
the $\eta$--Cristoffel connection and
$T$ is the $\eta$--torsion, \ie\ the shew--symmetric
(with respect $\eta$) part
$$T\eta= \half\big[g\partial g^{-1}
\eta-(g\partial g^{-1}\eta)^T\big].$$
Then $\CL_v= \CL_v^\eta+i(v)T+T\, i(v)$.
$\CL_v^\eta$ is the topological
RG flow. Since this flow is trivial ($\CL_v^\eta \Gamma=\Gamma$), we
have
$$\CL_v\Gamma= \Gamma + i(v)[T,\Gamma]+ [T,i(v)\Gamma]=[i(v)T,\Gamma
],$$
Then
\eqn\twoQ{Q=i(v)T= v^i g\partial_i
g^{-1}\Big|_{\eta-\rm skewsymmetric\ part}.  }
This equation also shows that $Q$ is hermitian
with respect to the metric
$g$ (thus the second equation in \RGQ\ follows from the first).
\twoQ\ should be compared with \isingQ .
It is manifest that the two results
agree provided $v= \sum_k w_k\, \partial_k$.
Then we generalize \shorthand\
by putting $C=i(v)\Gamma$. The matrix $C$ is
known as the `superpotential'.
In Appendix C of \topatop\ is also shown that (cf.
\otherconn\ and \isingQ )
\eqn\susyQ{T=\beta[Z,\Gamma]\qquad\Rightarrow\qquad Q=\beta[Z,C].  }
As in the Ising case, \susyQ\ allows writing the connection in
terms of $Q$ only, leading to a
simplification of the \ttstar\ equations.
For a more explicit connection with
Ising model see next subsection, where
we consider N=2 LG models.

The matrix $Q$ has also other physical
interpretations. Using the general N=2
Ward identity relating dilatations and axial
$U(1)$ rotations one shows \refs{\topatop \newindex}
$$Q_{kl}=-\half \langle l|Q^5|k\rangle,$$
where $Q^5$ is the (in general non--conserved) axial charge
$$Q^5=\oint \psibar \gamma_0\gamma^5\psi.$$
Whenever there is a conserved
$R$--charge,
the eigenvalues of the matrix $Q_{kl}$
agree with the values of the conserved
charge on the ground states, even if
the charge $Q^5$ itself {\it is not
conserved}\foot{The analog in the
Ising case is that $Q$ is equal to the
conserved angular momentum even if
we have rotational invariance around a
point distinct from the origin.}.
At criticality, there is always a conserved $R$--charge. The
charges $q_k$ of the Ramond ground states are related to the
dimension
of the primary chiral operators \chiralring\ as
$h_k=(q_k-q_0)/2$, where $q_0$ is the smallest $q_k$. In particular,
the
central charge is given by
$\hat c= 2|q_0|$. Then we can see the eigenvalues of
$Q$ as a generalization off-criticality of the
conformal
dimensions $h_k$ and central charge
$\hat c$. Indeed, these eigenvalues are
stationary at a conformal point
where they agree with the conformal charges
$q_k$. Thus $Q$ is an alternative to the Zamolodchikov's
$c$--function
\zamolo\ which has the advantage of being computable from the
\ttstar\
equations\foot{One can also show  that for a conformal
N=2 family $g/g_{1\bar 1}$
is equal to the Zamolodchikov metric \topatop.}.

There is another, more interesting, physical interpretation of
$Q$. Indeed $Q$ is a new  susy index
\newindex  . To see this, let us
quantize the N=2 model on a line instead of a
circle. We make the genericity
assumption that the theory has a mass--gap and
multiple vacua. Then
the Hilbert space contains solitonic
sectors interpolating between two {\it
distinct} vacuum configurations, $k$ and $l$, at $x=\pm\infty$
respectively \FMVW.
 For concreteness,
we define the theory in a segment of length $L$ and  take
the thermodynamical limit at the end. Then  a `modular
transformation'
shows \newindex\
$$Q_{kl}=\lim_{L\rightarrow\infty} {i\beta\over 2L} \Tr_{(k,l)}
\left[(-1)^F F
e^{-\beta  H}\right]$$
where $F$ is the (conserved) Fermi charge
and $\Tr_{(k,l)}$ denotes the trace
over the solitonic sector specified by
the boundary conditions $(k,l)$. $Q$ is an
index in the sense that it is independent
of the D--term. This is also the
reason why it is computable.
This `thermodynamical' quantity
receives non--trivial contributions from
multi--soliton sectors.
Then from $Q$ we can extract exact
information about the mass spectrum,
the soliton interactions, the Fermi number
fractionalization, the RG properties  of the
model, etc.

\subsec{The Canonical Coordinates}

Comparing sects. 2.2 and 3.1
 we see that the \ttstar\ equations for
a family of N=2 models have the same form as the Ising ones
\diffeqns\ provided {\it i)} all
fields $\phi_i$ have $F=0$, and {\it ii)}
we can
find new coordinates $w_k(t_i)$ in
which the ring coefficients take the form
$C_{ij}^{\ k}=\delta_{ij}\delta_j^{\ k}$.
The first condition is automatically
satisfied for unorbifoldized LG models.
On the other hand, it is a basic geometric consequence of
the TFT axioms that such {\it canonical coordinates}
exists under the additional
assumption that no element of $\CR$ is nilpotent
\refs{\cancoordinates\ ,\solttstar}.
Physically, this means we have
 a mass--gap\foot{However, the presence of a mass--gap
may be a slightly stronger
requirement than the semi--simplicity of $\CR$.}.

Since we are primarily interested
in the {\it physics} of this correspondence,
we begin by discussing  a concrete class
of models \ie\ the Landau--Ginzburg
models (LG) \landgins\ with a polynomial
superpotential $W(X)$. For these
theories there is a direct physical
identification with the Ising model,
explained in sect. 5 below.

\vskip 5mm
\noindent{\it The family of the $A_n$--minimal model}

Without loss of generality, we take a
superpotential of the form
\eqn\LGsuper{W(X;t_k)={1\over n+1}
X^{n+1}+t_{n-1}X^{n-1}+\dots+
t_1 X+t_0.  }
For generic $t_k$'s, $W(X;t_k)$ is
a Morse function\foot{I.e. all
critical values are non--singular
and all critical values are distinct.}.
We assume we are in this generic situation. As (local)
coordinates in the family we take the
critical values $w_i\equiv W(X_i;t_k)$
where $X_i$ are the critical points
 of $W(X)$ defined by
$$W^{\prime}(X_i;t_k)=0.$$

We work in the {\it canonical}
basis for the ground states.
This is the `point basis' of \topatop\ but
normalized such that $\eta=1$.
An element $|j\rangle$  of this basis belongs to
the topological class $|f_j(X)\rangle$,
where the $f_j(X)$'s are polynomials
such that\foot{This condition fixes
$f_j(X)\in\CR$ uniquely.}
$$f_j(X_i)=\delta_i^j
\sqrt{W^{\prime\prime}(X_i)}.$$
The canonical basis diagonalizes $\CR$
$$g(X) f_j(X) = g(X_j) f_j(X) \qquad {\rm in}\ \CR.$$

Let us compute the action of the observable
$$\CO_i\equiv{\partial W(X;t_k)\over \partial w_i},$$
in the canonical basis.
Using that
$W^{\prime}(X_i;t_k)=0$ in $\CR$, we get
\eqn\whatC{\CO_i|j\rangle\equiv\big(C_i\big)_j^{\
k}|k\rangle=\delta_{ij}\delta_j^{\ k}|k\rangle,}
\ie\ the critical values $w_k$ are just
the canonical coordinates we look for.
Adding a constant to the superpotential $W(X)$ does not change the
physics. Then from
$\sum_j {\partial\over \partial w_j}= n{\partial \over \partial
t_0}$,  it follows that
\eqn\euclinv{\eqalign{&\sum_j{\partial g\over \partial w_j}=0\cr
&\sum_j\left( w_j{\partial g\over \partial w_j} -\bar w_j{\partial
g\over
\partial \bar w_j}\right)=0,\cr}  }
where the second equation follows from the independence of $g$ from
the
overall phase of the superpotential\foot{Indeed
this phase can be absorbed by a
redefinition of the Fermi fields.}.
The N=2 Ward Identities \euclinv\ just express the fact that
$g(w_1,\dots, w_n)$ is Euclidean invariant in the $W$--plane, \ie\
that our
\ttstar\ metric has the symmetry properties required for the
$n$--point
function of a $2d$ QFT!
Then $g$ can be written in terms of
the Euclidean invariants of the points
$w_i$.
These invariants are objects of direct
physical interest. They are the lengths $|w_i-w_j|$
of the segments\foot{These lengths are
positive since $W(X)$ is assumed to be
Morse.} connecting two points
and the angles between these segments.
The first invariant is just one--half the
mass of the soliton interpolating
between the two vacua $i$ and $j$ \FMVW , and
the second is related to the $S$ matrix
for the scattering of the corresponding solitons \newindex .
There is a subtle
reason for this unexpected symmetry.
The \ttstar\ theory leads to the identification
\eqn\identification{\left({1-g(w_1,\dots,w_n)\over
1+g(w_1,\dots,w_n)}\right)_{jk}\leftrightarrow i
{\langle\mu(w_1)\dots
\sigma(w_j)\dots\sigma(w_k)\dots\mu(w_n)\rangle\over
\langle\mu(w_1)\dots\mu(w_n)\rangle},  }
from which the Euclidean invariance is manifest.

The identification \identification\
should be taken with a grain of salt. In
fact we have shown that the equations
 are the same, but this does not mean that
the {\it solutions} are equal for the
 two theories. I.e. in general they
satisfy different boundary conditions.
Selecting the correct boundary condition
is a subtler point. The explicit connection
between the boundary conditions
satisfied by the Ising correlations and the
metric $g$ for \LGsuper\ will be
discussed in sect. 4 (for special cases).

\vskip 3mm
\noindent{\it The general case}

The general case is very similar. The canonical
coordinates $w_i$ always exist.  The way to see this is to note
that in the sector connecting vacua $i$ and $j$, the supersymmetry
algebra has a central term $\{ Q^+,\bar Q^+\} =\Delta_{ij}$,
satisfying $\Delta_{ij}=-\Delta_{ji}$.  Moreover by additivity
of the central term, one sees that $\Delta_{ik}=\Delta_{ij}+
\Delta_{jk}$, from this one deduces that we can choose $w_i$
for each critical point, which are unique up to an overall shift
and such that $\Delta_{ij}=w_i-w_j$.  The existence of canonical
coordinates $w_i$ can also be shown in a purely topological
setting \refs{\cancoordinates\ ,\solttstar}.
These coordinates have all the properties needed in the above
discussion (in particular in this basis
topological metric $\eta$ is always diagonal).
 The mass of
the soliton connecting the vacua $i$
and $j$ is expressed in terms of the
canonical coordinates as
\eqn\bogomolnyB{m_{ij}= 2|w_i-w_j|,  }
and the angles between the segments
connecting the `points' $w_i$ are
still connected with ratios of $S$--matrix elements.
This follows from the IR
asymptotics of the solution to the \ttstar\
equations and the thermodynamical
interpretation of the $Q$ matrix \newindex.
In fact comparing with the Bogomolnyi bound
of \FMVW\ we see that the $w_j$ are still
the critical values of the
superpotential (even for non--LG models).
Then the analysis above carries over word for word
to the general case.

There is however one major difference.
In the general case it
is no longer obvious that
the solutions to the \ttstar\ equations
should be regular everywhere in the
`$W$--plane'. Physical intuition certainly
requires regularity in the subspace
corresponding to the relevant or
marginal perturbations, but the
irrelevant (\ie\ non--renormalizable)
ones are more tricky and possibly ill--defined.

\vskip 5pt
\noindent{\it Fancier Points of View}
\vskip 2pt

At a more speculative level one
would like to argue the other way around.
That is, one starts from the `magical'
euclidean invariance \euclinv\
to construct a $2d$ QFT over coupling space.
The correlations of this
`target' theory will compute non--trivial
quantities for the original N=2 theories.

In principle this can be done using
the Euclidean reconstruction theorem
\axioms. This theorem allows to
reconstruct a QFT out of a set of would--be
$n$--point functions provided they
satisfy the Osterwalder--Schrader axioms
OS1,---, OS5 \axioms. Take any physical quantity
$A$ for the model \LGsuper,
and consider it as a function of the couplings
$(w_1,\dots,w_n)$. By construction
the would--be $n$ points functions
$A(w_1,\dots,w_n)$ satisfy two of the basic
axioms, namely OS2 {\it (euclidean
invariance)} and OS4 {\it (symmetry)}.
In fact\foot{This discussion is valid if $A(w_1,\dots, w_n)$
is a univalued function. This needs not to be so, since the change of
variables $t_i\leftrightarrow w_j$ can be done only locally.
If $A(w_i)$ is multi--valued a standard analysis shows that the
would--be
fields are anyons rather than bosons. For the LG case the
corresponding
`statistics' can be fully described in terms of Picard--Leschetz
theory \singT.}
, $A(w_1,\dots,w_n)$ is
invariant under permutations of the $w_k$'s,
just because it is a function
of the original couplings $t_j$,
which are themselves symmetric functions
of the $w_k$.

Whether the other three axioms are satisfied depend on the quantity
$A$. These should be seen as restrictions on the objects $A$ which
can be computed by the reconstruction program.
It is easy to construct quantities $A$ which satisfy
OS5 {\it (cluster property)}. This requires
$$A(w_1,\dots, w_r,
w^\prime_1,\dots,w^\prime_s)\Big|_{|w_i-w^\prime_j|\rightarrow\infty}
\sim A(w_1,\dots, w_r)\, A(w^\prime_1,\dots, w^\prime_s).$$
Indeed each critical value
corresponds to a classical vacuum. As the two groups of
critical values are taken
apart the potential barrier between the corresponding vacua grows.
There is a rigorous bound on the corresponding tunnelling amplitude,
$${\rm tunnelling\ amplitude}=
O\big(\exp[-2\beta |w_i-w^\prime_j|]\big).$$
In the limit
$|w_i-w^\prime_j|\rightarrow\infty$
the sectors of the Hilbert space
built over the two sets of vacua decouple.
Thus, say, quantities\foot{Here
$\langle\cdots\rangle_{ij}$ represents
the expectation value in the soliton
sector connecting the $i$--th vacuum
to the $j$--th one.  In this sector $W(X)\rightarrow w_i$ (resp.
$w_j$)
for $X\rightarrow -\infty$ (resp. $+\infty$).} like
$\exp[\langle\CO\rangle]$ or
$\det[\langle\CO\rangle_{ij}]$
satisfy OS5.
The axioms OS1 {\it (regularity)}
is  not really difficult to realize. In particular the reality
condition looks very natural: It requires $A$ to be real whenever
the coefficients of $W$ are.  The axiom which is non--trivial
 is OS3 {\it (reflection positivity)}.
This is the essential condition on the quantity $A$ needed to
compute it by reconstructing a QFT in coupling space.

At least for some N=2 models,
our correspondence with the Ising problem
can be seen as a way
to reconstruct a `target' QFT (\ie\ our
old friend the Ising model).
  In general, it may be necessary
to go to a covering of the Euclidean plane in order to realize
the `target' QFT, as the fields may have non-abelian anyonic
statistics.  It may also be that there are other quantities $A$
allowing reconstruction beyond those discussed
in the present paper.

\newsec{The Physical Link Between the Ising Model and N=2 Susy}

We have seen that the Ising
correlations and the $g$ metric for a
`generic' class of N=2 models lead
to the same abstract mathematical problem
described by the \ttstar\ geometry.
It is natural to ask whether there
is a more direct physical link between N=2
 susy and the Ising model. Besides its
intrinsic interest, this question is very
 important for practical reasons. In
fact, although these two different area of physics are governed by
the
same differential equations, the {\it solutions} are, in general,
different  for the two cases. \ie\ the universality at the level of
equations breaks down at the level of the boundary conditions.
 Having a direct link, we can relate the actual
solutions of the two problems.
This would be nice since we know the explicit
solutions of \ttstar\ corresponding
  to the Ising correlations. To
compute, say, the $\tau$ function
$\langle\mu(w_1)\dots\mu(w_n)\rangle$, we can
insert a complete set of states.
Since the theory is free, these are just $k$
fermions of momenta $p_i$ ($i=1,\dots,k$).
In this way we can construct all
correlation functions in terms of the form factors
\eqn\formfact{\langle p_1,\dots,
p_k|\mu|p^\prime_1,\dots,p^\prime_h\rangle.  }
These functions are known \refs{\isingFF\french}.

Then a direct link would allow to compute
 $g$, the index $Q$, etc. in terms of
the form factors for the Ising model.
It may seem that this procedure can work
only for very special models, since
it is quite rare that the boundary
condition satisfied by ground--state
 metric is the one corresponding to
the Ising correlations.   We shall see first of all
that there {\it does exist} an $N=2$ theory for each $n$-point
correlation of Ising model, which satisfies exactly
the same boundary conditions (these corresponds to theories
where the chiral field lives on a hyperelliptic Riemann surface).
Moreover,
as we shall see, the class of N=2 systems that can be related to
the Ising model is much bigger. The
idea is that we can `twist' the fermi field
of the Ising model by inserting in
the correlations other operators besides the
order/disorder ones. This changes the
boundary conditions for the fermion and
consequently the boundary conditions
for the \ttstar\ equations. In this way we
can generate families of solutions to \ttstar\  in
terms of the Ising form factors.  It is not clear whether or not
in this way we get the full boundary conditions allowed by
regularity of \ttstar\ equation.  At any rate,
a more general approach to solving
the \ttstar\ equations, which leads
to a classification program for $N=2$ quantum field
theories in two dimensions will be presented elsewhere
\ref\cvta{S. Cecotti and C. Vafa, in preparation.}.

\subsec{SQM and the Ising Model }

The most direct way to make contact between
 N=2 Landau--Ginzburg and the Ising
model is by looking at the $1d$
version of the LG model, \ie\ by going to SQM.

 For a non--degenerate one dimensional
 LG model the $Q^+$--cohomology  is
isomorphic (as a ring) to that of the
corresponding $2d$ model, and the ground
state metric
 is the same \compact. Of course, in the SQM case we
can also compute the metric from the overlap integrals for
 vacuum wave function (as we did in sect.(4.2)).

As always, we identify the SQM Hilbert
space with the space of square summable
forms through the identification \asforms\
$$\psi^i\mapsto dX^i,\qquad \psi^{\bar i}\mapsto d\bar X^i.$$
The  two supercharges
having fermi number $F= +1$,
$Q^{+}$ and $\bar Q^{-}$, act  on forms as \compact\
\eqn\whatQp{Q^+\alpha=\bar\partial\alpha + dW\wedge\alpha,\qquad
\bar Q^-\alpha=\partial\alpha + d\overline W\wedge\alpha,  }
where $W(X_i)$ is the superpotential.
As in sect.2, we denote by $\Lambda$ the
contraction with respect to
the \"Kahler form defined by the $D$--term. Then
$$(Q^+)^\dagger=i[\Lambda, \bar Q^-],
\qquad (\bar Q^-)^\dagger=-i[\Lambda,
Q^+].$$
The form $\varpi_j$ associated to the wave--function of a susy
ground--state satisfies,
$$Q^+\varpi_j=\bar Q^-\varpi_j=\Lambda\varpi_j=0.$$

We focus on the case in which we have
a LG model with only one superfield $X$.
Then consider the holomorphic change of variables
\eqn\changevar{X\mapsto w\equiv W(X). }
Comparing \defcha\ with \whatQp\ we see that ($m=1$)
$$Q^+= W^\ast\bdi,\qquad \bar Q^-=W^\ast\CD.$$
Then the map \changevar\ transforms
the LG model into the Ising model in the
$W$--plane. Stated differently, \changevar\ maps the
Schroedinger equation for a zero--energy
state of the $1d$ LG model in $X$--plane into the
Euclidean Dirac equation in the $W$--plane \dirac.
Thus, all we said in sect.(2.2) for the Ising model
applies word--for--word to the $1d$
LG model. In particular,
the wave function viewed on the $W$--plane, which we denote by
$W_\ast\varpi_j$, can be written as $W_\ast\varpi_j=i\bdi\psi_{j+}$,
where $\psi_{j+}$ is singular only at the branching points $w_i$, and
these
singularities encode  the $Q^+$--class of the vacuum $\varpi_j$.

However on $W$--plane $W_\ast\varpi_i$ is not univalued. $\varpi_i$
is required
to be
univalued only in the $X$--plane. The branching
points of $W_\ast\varpi_i$ are just
the critical values $w_k$ of $W(X)$.

In general not all the pre--images of
a given critical value $w_i$ are critical
points. If all the pre--images of
any critical value are critical points  we
 say that the superpotential $W(X)$
is {\it nice}.  If $W(X)$ is both `nice' and
Morse life is particularly easy.
In this case the regularity of
$\varpi_j$ in $X$--space requires
\eqn\diracope{\psi_{j+}(w)\sim {a_{jk}\over (w-w_k)^{1/2}}+\dots\quad
{\rm as}\ w\rightarrow w_k.  }
Using standard QM techniques, one
easily sees that, as $w\rightarrow\infty$,
$\psi_{j+}=O(\exp[-2|w|])$.
Thus, for $W(X)$ `nice' and Morse,
$W_\ast \varpi_j$ not only satisfies  the
same equation \dirac\ as the Ising
wave functions \wavefunction , but also
the same boundary conditions \diracope .
Thus the SQM wave--functions
are just the correlation functions of
the Ising model.
$$W_\ast\varpi_j(w)=i\bdi{\langle\psi_+(w)\mu(w_1)\dots
\sigma(w_j)\dots\mu(w_n)\rangle
\over \langle\mu(w_1)\dots\mu(w_n)\rangle}.$$

This explains why the Ising correlations
satisfy the \ttstar\ equations: The
Ising Model {\it is} an N=2 LG model
for some `nice' superpotential!

Conversely, for `nice' Morse superpotentials
we can write the explicit
solutions to the Schroedinger equation
in terms of Ising form factors.
The `twisting' procedure mentioned at
the beginning allows to extend the space
of models which can be solved in terms
 of Ising form factors. A further
extension can be obtained considering
more general ${\bf Z}_m$ twist--fields
for the theory of massive free fermions.
This allows to solve simple models
without a mass--gap.  In the next subsection
we present some examples where \ttstar\ equations
are solved through the correlation functions of Ising
model.

\subsec{Solving the \ttstar\ Equations by the Ising Map (Elementary
Case)}

In this subsection some example of models for which the actual $g$
can be
written in terms of the Ising form factors is discussed.  In
particular
we discuss the cases of $N=2$ sine-Gordon model, the Chebyshev
superpotentials, and the hyperelliptic configuration space for
chiral fields (which is the exact mirror of Ising model in the
$N=2$ set up).

\noindent\underbar{\it N=2 sine--Gordon}

Let
\eqn\sinegordon{W(X)=\lambda \cos(X). }
The critical values are $\pm \lambda$
and all points with $W(X)=\pm \lambda$
are critical. So \sinegordon\ is `nice'
but it fails to be Morse since the
critical values are not all distinct.
To get a Morse function one has just to
make the identification $X\sim X+2\pi$,
\ie\ let $X$ take value on a cylinder
$\CC$. Then the vacua of the original
model correspond to the $Q^+$--cohomology
with coefficients in the flat bundles
$V_\chi$, where $\chi$ are the unitary
representations of $\pi_1(\CC)=\bf Z$.
 These are just the Block waves, \ie\ the $\theta$--vacua, such that
$$T|\theta; a\rangle =
 e^{i\theta}|\theta;a\rangle \qquad 0\leq \theta < 2\pi,$$
where $T$ is the operator which translates $X$ by $2\pi$.
 For each $\theta$ there are two
vacua corresponding to the two critical values $\pm\lambda$.

We start with the simpler
case\foot{In fact, $\theta=\pi$ is even simpler.
The wave--functions are easily
expressed in terms of Bessel functions. Up to
normalization, the two $\theta=\pi$ vacua are given by
$$\varpi_{\pm\lambda}(\theta=\pi)=i\bdi\Big\{\zeta_0[2(W(x)\mp
\lambda]-\zeta_1[2(W(x)\mp \lambda]\Big\},$$
where the functions $\zeta_l(z)$ are as in Eq.\bessel .}, \ie\
 $\theta=0$. This
corresponds to wave--functions univalued on the cylinder.
{}From the above discussion, we see
that --- after the change of variables
\changevar\ --- the wave functions
for the two $\theta=0$ vacua are just the
Ising functions
$$\varpi_{\pm\lambda}(w;\theta=0)=i\bdi {\langle \psi_+(w)\sigma
(\pm\lambda)\mu(\mp\lambda)\rangle \over
\langle\mu(\pm\lambda)\mu(\mp\lambda)\rangle}.$$

 So, in this case the boundary condition
for the metric $g$ agrees with that
for the Ising two--point function.
In the canonical basis one has
$$g(\lambda;\theta=0)=\exp[\sigma_2 u(4|\lambda|)].$$
In term of $u(z)$,  \ttstar\ equations for $\theta=0$ reduce to
radial
sinh--Gordon (a special case of PIII \piiimath)
\eqn\spPiii{ {\partial \phantom{b}\over
\partial z}\left( z{\partial u(z)\over \partial
z}\right)= {z\over 2} \sinh(2u). }
{}From the theory of the PIII equation
\piiimath\ we know that there is one
regular solution for each value of
the boundary datum $r$, $|r|\leq 1$, where
$$u(z)\sim r\log z \qquad {\rm as}\ z\rightarrow 0.$$
Recalling \twoQ , we see that the eigenvalues of the $Q$ matrix are
$$\pm \half |\lambda|{\partial u(4|\lambda|)
\over d|\lambda|} \sim \pm \half |r|,
\quad {\rm as}\ \lambda\rightarrow 0.$$
In the Ising picture, $Q$ is the mean
angular momentum. As $\lambda\rightarrow
0$, using the Ising operator expansion,
we can replace the order/disorder
operators by a single $\Psi$ at the origin.
Then in this limit we recover
invariance under rotations around the origin.
The angular
momentum of $\Psi$ is $\pm 1/2$. Thus, the boundary
condition associated to the Ising
2--point function  is $r=\pm1$.

Next, we construct the wave--functions
for a general value of $\theta$.
 When $X$ goes around
the cylinder the wave function picks up a phase $\exp(i\theta)$.
 Since $\Psi$ is now multivalued, we
cut the cylinder along one generator, ${\rm Re}\, X=a$. When
 crossing this cut, $\Psi$ gets
multiplied by the above phase. The image in the $W$--plane of
this cut is a branch of hyperbola
$\cal Y$. In the Ising language, we mimick this by
inserting an operator on this curve
which when crossed by the free fermion produces the right phase.
 However this is not quite correct, since
the pre--image of this curve corresponds to two
generators of the cylinder, ${\rm Re}\, X=a$ and ${\rm
Re}\,X=a+\pi$. Then it is more convenient to make {\it two} cuts
along
these two generators, in such a way that
at each cut the wave--function picks up a phase
$\exp[i\theta/2]$.
The images in the $W$--plane of
these two cuts correspond to the same curve
$\cal Y$, but with opposite orientations.

 Since the fermion is free, we make
it complex by adding a spectator imaginary part, which
does not couple to spins fields $\sigma$ and
$\mu$. Let $J_{\mu}$ be the corresponding $U(1)$ current
and insert in the above correlation functions the operator
$$\exp\left[\pm i{\theta \over 2}\int_{\cal
Y}\epsilon_{\mu\nu}J^{\mu}dw^{\nu}\right]$$
where $\pm$ corresponds to the two different sheets. This operator is
nothing else than $exp[-i\theta F/2]$ where $F$ is the `target' Fermi
number.

For these `twisted' Ising correlation functions we can repeat the
analysis of sect.2, getting the same differential equation. Again,
$$g(|\lambda|,\theta)=\exp[\sigma_2 u(4|\lambda|,\theta)],$$
with $u(z,\theta)$ the solution to Eq.\spPiii\ with
\eqn\generalsol{{\exp[u(4|\lambda|,\theta)]-1\over
\exp[u(4|\lambda|,\theta)]+1}=
{\langle 0| \sigma(\lambda) \exp[\pm i\theta F/2]
\sigma(\lambda)|0\rangle\over \langle 0|\mu(\lambda) \exp[\pm i\theta
F/2]
\mu(-\lambda)|0\rangle}. }
Now, the crucial fact is that, although
the wave--function is multivalued in the
$W$--plane, the correlation functions in
 the rhs of \generalsol\ are not. In
particular the insertion of the two
operators $\exp[\pm i\theta F/2]$
(corresponding to the two sheets)
have the same effect (since only the real
part of the fermion couples to $\sigma$
and $\mu$). So \generalsol\ is
unambiguously defined.

Let us compute the corresponding
boundary datum $r$, equal to twice the
angular momentum at $\lambda=0$.
Going around the cylinder $\CC$
we make a turn
around the origin and the wave
function picks up a phase $\exp[\pm i\theta]$.
Hence the orbital angular momentum
$m$ is equal to $\pm \theta/2\pi$ mod.1.
Adding the spin--part, we get
$$r(\theta)=2l=\pm \left(1-{\theta\over \pi}\right).$$
Notice that all $|r|\leq 1$ appear,
and hence all regular solutions to PIII can
be constructed this way. Since the
\ttstar\ equations for all the `massive' N=2
model with Witten index 2 can be recast
in the form \spPiii , this justify our
claim that for all such models the metric
$g$ can be written in terms of Ising
form factors.

Let us give a more explicit formula for the
 rhs of Eq.\generalsol .
Since the Ising model is just a free fermion
$$\eqalign{\langle 0| \sigma(\lambda)
e^{i\theta F/2}\sigma(-\lambda)| 0\rangle
&= \sum_{k=0}^{\infty} {1\over k!} {1\over 2^k}\sum_{e_i=\pm 1}
 \exp\Big[i\half \sum_{i=1}^k e_i\theta\Big]
 \int \prod^k_{i=1}
{dp_i\over 4\pi \sqrt{p_i^2+m^2}} \times \cr
&\quad \times \langle
0|\sigma(\lambda)|p_1,\dots,p_k\rangle \langle p_1,\dots,
p_k|\sigma(-\lambda)|0\rangle,\cr}$$
where we sum over the two possible $U(1)$
charges of each free fermion in the intermediate state.
A similar representation holds for the $\mu$ 2--point function.

Putting everything together, we get
$$u(z,\theta)=\sum_n [\cos(\theta/2)]^n u_n(z),$$
where $u_n(z)$ is the $n$--intermediate fermion
contribution to the Ising answer; $u_n(z)$ can be
extracted from the known form factors. As
 $\theta\rightarrow \theta+2\pi$, $u(z,\theta)$ changes
sign. Then
only odd $n$'s contribute. In particular
$$u(z,2\pi-\theta)=-u(z,\theta).$$
Comparing with \piiimath\ we get
$$\eqalign{u_{2n+1}(z)=& {2\over 2n+1}
\int \prod_{i=1}^{2n+1}{e^{-z\cosh
\theta_i}\over \cosh\left({\theta_i-
\theta_{i+1}\over 2}\right)}{d\theta_i \over
4\pi}\cr &({\rm here}\ \theta_{2n+1}\equiv \theta_1).\cr}$$
Additional information about these
solutions can be found in \topatop\ and
\newindex .  Of course, we can also write the
$\theta$--vacuum wave--functions in
terms of Ising form factors in a similar guise.
It is remarkable that we can
solve the N=2 sine--Gordon Schroedinger equation by the Ising map.

\vskip 5mm
\noindent\underbar{\it Chebyshev Superpotentials}

We can use the Ising map to compute
the ground--state metric for Chebyshev
polynomial superpotentials
\eqn\chebyshev{W(X)=\lambda T_n(X),
\qquad {\rm where}\quad T_n(\cos x)=\cos(n x). }
The change of variable $f_n\colon X\mapsto
 \cos[Y/n]$ maps this model into the
sine--Gordon one. As we saw in sect.(4.1),
the susy charges (and hence the
zero--energy Schroedinger equation)
transform functorially under such change of
coordinates. If we
look at $Y$ as a variable taking value
in the cylinder $\CC$ as above, then the
Chebyshev vacua can be seen as solutions
to the sine--Gordon belonging to
certain representations $\chi_i$
($i=1,\dots, n-1$) of $\pi_1(\CC)$. Since
$Y\rightarrow Y+2\pi n$ acts as the
identity on $X$, these representations
should be direct sums of those with
$\theta=2\pi k/n$. The vacua with
$\theta=0$, $2\pi$, cannot appear
because they would not be singularity free in
$X$ space. So $k=1,\dots, n-1$.
Moreover, the Chebyshev vacua are
required to be odd under the symmetry
$X\leftrightarrow -X$. These two requirements
give $n-1$ vacua. Since this is
the Witten index of the model \chebyshev,
they fix completely the vacuum wave
functions and hence the metric $g$.

Then $g$ can be written in terms of
$u(z,2\pi k/n).$ Details can be
found in \topatop.

\vskip 5mm
\noindent\underbar{\it Other Models}

The N=2 sine--Gordon is the first
model in a family of `nice' superpotentials
having Witten indices $n=2,3,\dots$,
whose ground--state metric not only satisfies
the same equations as the corresponding
Ising correlation functions, but also the
same boundary conditions. To understand
the general case, notice that the
important property of $W$ which made the sine--Gordon case
work was
$$\left({dW\over dx}\right)^2=P_2(W),$$
where $P_2(\cdot)$ is a polynomial of degree 2.
{}From this equation it is obvious
that $w$ is a critical value if and
only if is a root of $P_2(\cdot)$.
Conversely,
all inverse images are critical points. Then the
above differential equation guarantees that $W$ is `nice'.
This argument is easily generalized.
The next model satisfies the equation
$$\left({dW\over dx}\right)^2=P_3(W),$$
where now $P_3(\cdot)$ is a cubic polynomial.
The solution is $W=\wp(X)$,
and the argument above shows that this potential
is also `nice'. To make it Morse,
one has just to identify
\eqn\identifyX{X\sim X+n\omega_1+m\omega_2,  }
where $\omega_i$ are the periods of the elliptic curve
$\CC$ defined by $\wp(X)$.

We will now describe this result more invariantly in a more
general setup.
This turns out to be useful in understanding
its relation with the Ising model. Consider a genus $g$ curve
(Riemann surface)
$\CC_g$ as a (branched)
double cover of $\BP^1$, i.e., a hyperelliptic curve. This cover
is described by a
degree two meromorphic function
$$W\colon \CC_g\rightarrow
\BP^1.$$
whose critical values are the branching points $w_1$, $w_2$, ...,
$w_n$
(the other $4g-n$ branching points are set to infinity). Clearly, all
inverse
images of
these points are critical for $W$. Then we take $W$ as our `nice'
superpotential provided we take as
field ($P\in \CC_g$)
\eqn\whatfield{X(P)=\int_{P_0}^P\omega,}
where $\omega$ is a holomorphic differential\foot{More concretely,
if the hyperelliptic curve has equation
$$y^2=\prod\nolimits_i (W-e_i),$$
we take $\omega=dW/y$. Then we have $(W^\prime)^2=\prod_i(W-e_i)$,
where a prime denotes the derivative with respect the local parameter
$X$.} over $\CC$.
($X$ is well defined up to  periods of $\omega$).  LG
models with these kind of superpotentials as well as those
associated to more general Riemann surfaces have been introduced
and studied in detail by B. Dubrovin, Ref. \highermod (see
also \cancoordinates ).

The ground--state metric $g$ for these
models satisfy the Ising correlation
equations. Indeed, the $w_k$'s are again
canonical coordinates \cancoordinates.
Moreover, they satisfy the same boundary
conditions. This can be seen by
comparing the corresponding SQM
wave functions with the Ising ones.
In fact the Ising correlations are
uni--valued when continued on the
hyperelliptic surface $\CC_g$ just as the wave functions
of the $LG$ theory is\foot{
Note that by the above reasoning the two point function in the Ising
model
can also be related to an $N=2$ superpotential
$W=(X^2-1)^{1/2}$ defined on the double cover of the plane.}.  The
two
wave-functions are indeed
the same.

Thus the reconstruction program
outlined in sect. 3.2  can be carried out for
this class of superpotentials.
The target QFT we get this way is just
the Ising model itself.

\newsec{Geometry of the N=2 $\tau$--Function}

In sect.3 we have seen that the \ttstar\ equations for
a `massive' N=2 theory
take the same form as the equations
for the Ising correlators. There we have
discussed the  N=2 meaning of all Ising quantities but for one:
the $\tau$ function.
One of the purposes of the present paper is to elucidate
the physical meaning of the $\tau$
function from the N=2 viewpoint. As a
preparation to the more substantial
physical analysis of sect. 6, we begin with a general
discussion on its geometrical origin in the framework of
topological anti--topological fusion (\ttstar) \topatop .
For concreteness we fix
our attention to the LG case,
but our conclusions are fairly general.

\subsec{K\"ahler metrics in \ttstar}

{}From many points of view, \ttstar\
is a generalization of the so--called
`special geometry' (variations of
Hodge structure \hodge). However, as a
generalization of special geometry, \ttstar\
is disappointing in one respect: The
metric $g_{i\bar j}$ is not  K\"ahler
in general, and even for a critical
theory it is (conformally) K\"ahler only after
restriction to marginal directions.

Luckily in \ttstar\ there is another natural
(hermitian) metric for $\cal R$
which {\it is}
 K\"ahlerian. It is defined
by the following procedure.
Recall that the metric $g_{i\bar j}$ was defined in \topatop\
by considering an infinitely elongated sphere, in which we consider
topological twisting on the left-hemisphere, and anti-topological
twisting on the right-hemisphere, and we insert the operator
$\phi_i$ on the left hemisphere and $\bar \phi_j$ on the right one.
As mentioned in that paper, it is also possible to do a similar
thing on an arbitrary genus Riemann surface as long as the regions
in which we do the topological twisting and the ones where we
do anti-topological twisting are separated by infinitely long tubes.
Indeed the computation of the topological OPE, $C_{ij}^{\ k}$ and the
metrics
$g_{i\bar j}$ and $\eta_{ij}$ on the sphere
is sufficient to enable us to do an arbitrary such
computation on higher genus by the sewing techniques.  It turns out
the case of interest for us is the metric defined via a torus.

So let us consider a torus with the fields $\phi_i$ and $\bar \phi_j$
inserted on the left and right side of a flat torus
respectively which are infinitely
separated by two long tubes each of perimeter $\beta$
(sometimes we set $\beta =1$ by a choice of units).
We denote the resulting correlation by
\eqn\defK{K_{i\bar j}=\langle \phi_i \bar \phi_j \rangle_{torus}. }
As in sect.(3.1), we choose our
coordinates in coupling constant space, $t^i$,
such that  $\phi_i=\partial_i W$ and $\partial_0W=1$.
As we shall show below,
$ds^2=K_{i\bar j} dt^i d\bar t^j$
is a K\"ahler metric, i.e., (locally) there is a real function $K$
such that
$$K_{i\bar j}=\partial_i\partial_{\overline{j}} K.$$

{}From the previous sections we know that
\ttstar\  associates a $\tau$
function to each N=2 model. (We take Eq.\taudiff\ as definition of
the
$\tau$ function even for models which have
no canonical coordinates, \eg\ the
critical ones). It
turns out that $K$ is simply related to this
 $\tau$ function
\eqn\tauK{K=-  \log\tilde\tau+\tr(CC^\dagger) +f(t_i) +f^*(\bar t_i).
}
for some holomorphic function $f$ of moduli.
This equation is shown in sect. 5.2 below by showing
that $\partial \bar \partial$ of both sides of it are equal.

Then the $\tau$ function is the
analog  in toroidal \ttstar\ of $\int\epsilon\wedge
\overline{\epsilon}$
in `special geometry' \specgeom\
($\epsilon$ is the holomorphic $(n,0)$ form).
In fact, $\tilde\tau$ is a kind of `partition function'.
The reason for the additional term $\tr (C C^\dagger)$
in \tauK\ is that $\tilde\tau$, being
translational invariant in $W$--plane, is independent of $t_0$,
whereas
$K$ is not. For if we take $\phi_i=1$ then $<\bar \phi_j>_{torus}$
is not zero and is equal to $\tr \ \bar C_j$\foot{From the definition
of $K$ we
have
$${\partial\phantom{m}\over\partial t_0}K=
\big(\eta^{ij}\partial_i\partial_j F\big)^\ast,$$
where $F$ is the topological prepotential (the rhs is written in
terms of
 flat coordinates \refs{\TFT \flatcoordinates\cancoordinates}) .}
The second term in \tauK\ restores translational invariance .
In this sense then $\log \tilde\tau$, which is a
minor modification of $K$, may be a more natural
candidate for a K\"ahler potential, as the perturbation by the
identity
operator in the superpotential does not change the theory at all,
and so $\log \tilde \tau$ defines a K\"ahler metric on the {\it
physical}
perturbation space of the theory.

Let us show that the $K_{i\bar j}$
metric is K\"ahler.  Using the operator representation
of the path-integral on the torus, and taking flat
metric on the torus we have\foot{As mentioned earlier
we are taking the chiral field all to have zero fermion number
thus no need for insertion of $(-1)^F$ on the rhs of the equation
above.},
\eqn\kttt{K_{i\bar j}=\lim_{L\rightarrow\infty}{\rm Tr}\Big[(-1)^F
\phi_i e^{-LH}\overline\phi_j e^{-LH}\Big]={\rm tr}\big(C_i\bar
C_{\bar j}\big)}
In particular, $K_{i\bar j}$ is known once $g_{i\bar j}$ is
known and the topological correlations are known.

Using \covholo\ and \otherttstar\
\eqn\kahler{\eqalign{&\partial_k K_{i\bar j}=\partial_k {\rm
tr}\big( C_i\bar C_{\bar j}\big)=\cr &= {\rm tr}\big[(D_k
C_i)\bar C_{\bar j}\big]= {\rm tr}\big[(D_i C_k)\bar C_{\bar j}\big]=
\partial_i K_{k\bar j},\cr} }
and so $K_{i\bar j}$ is K\"ahler\foot{If we wish
to get a normalized metric, \ie\
with $K_{0\bar 0}^\prime=1$, we have just to divide
$K_{i\bar j}$  by $K_{0\bar 0}=\Delta$, the Witten index. Of course,
this
does  not spoil the K\"ahler property.}.

Eq.\kahler\ has the following generalization.
Let $T$ be a torus with a
flat metric $h$. Then
\eqn\follgen{K_{i\bar j}(T,h)=\int_Td^2z
\langle\overline{\phi_j}(z)\phi_i (0)
\rangle_{(T,h)},  }
has the K\"ahler property for all $T$ and $h$.
However, for finite periods, $K_{i\bar j}(T,h)$ does
depend on the D--term, and so is not an
index--like quantity in the sense of
\newindex .

\subsec{Differential Geometry of $\tau$--functions}

It remains to show Eq.\tauK.
Let us introduce some differential forms over coupling constant
space
\eqn\deforms{\eqalign{&\kappa=
\tr\big(\Gamma\wedge\bar\Gamma\big), \qquad \bar
\varrho = i(v) \kappa, \cr  &\Omega={\rm
tr}\big(QR\big), \qquad \xi=
\half\tr(QT)\equiv \half \tr(Qg\partial
g^{-1}).\cr}   }
where $R\equiv -[\Gamma, \bar\Gamma]$
is the \ttstar\ curvature and $T$ is the
torsion $(1,0)$ form. Notice that
$\kappa$ is just the Kahler form of the
metric \defK. Then from \kttt ,
\eqn\kalpot{\kappa=\partial\bar\partial K.  }

These forms are related by the following identities,
\eqn\lemma{\eqalign{&\Omega=\kappa-
\partial\bar\varrho, \hskip 1.5cm ,d\xi=\Omega\cr &
\bar\varrho=\bar\partial\left[ {\rm tr}\big(C\bar
C\big)-\half\tr\big(Q^2\big)\right].\cr}  }
Indeed, we have
$$\eqalign{\Omega &=-\tr\big(Q[\Gamma,\bar
\Gamma])=-\tr\big([Q,\Gamma]\wedge\bar \Gamma)= \hskip 2.5cm
{\rm using\ \RGQ}\cr
&=\tr(\Gamma\wedge\bar\Gamma)-\CL_v\,
\tr(\Gamma\wedge\bar\Gamma)=\kappa-\partial
\, i(v)\kappa,\cr}$$
which shows the first equality in \lemma . In terms of $T$ the
\ttstar\
equations take the form
\eqn\bardelT{\eqalign{&\bar\partial T=-[\Gamma,\bar\Gamma],\cr
&\Rightarrow \quad\bar\partial Q=\bar\partial\,
i(v)T=i(v)[\Gamma,\bar\Gamma].\cr}  }
Then we have
\eqn\two{\bar\partial\,\tr(QT)=
\tr([C,\bar\Gamma]\, T)-\tr(Q[\Gamma,\bar\Gamma])=2\Omega,}
where we used the identity
$[Q,\Gamma]=[T,C]$ which follows from \secondtt .

 On
the other hand, we have $DT=T\wedge T$
and $[\Gamma,T]=0$ (also a consequence
of \secondtt). Therefore,
\eqn\three{\partial\,\tr(QT)=
\tr([\Gamma,\bar C]T)+v^i\,\tr(T_iT_jT_k)dt^j\wedge dt^k=0.   }
\two\ and \three\ give the second equality in \lemma .

Finally, \bardelT\ yields
$$\eqalign{\bar\partial\, \tr Q^2
& =2\, \tr\big(Q \bar\partial Q\big)=
2\, \tr\big(Q[C,\bar \Gamma]\big)\cr
&= 2\,\tr\big(C [\bar\Gamma,Q]\big)=
-2\,\tr (C\bar\Gamma)+2\,\bar\CL_{\bar
v}\,\tr(C\bar\Gamma)\cr
&=-2\bar\varrho+2\bar\partial\,\tr(C\bar
C).\cr}$$
which proves the last equality in \lemma .
Comparing \taudiff\ with \deforms\ and using \lemma, we get
\eqn\holq{\eqalign{\partial
\bar\partial\log\tilde\tau &=
\partial\big[\bar\xi-\bar\varrho
+\bar\partial\tr(CC^\dagger)\big]=\cr
&=-\Omega-\partial\bar\varrho
+\partial\bar\partial\,\tr(CC^\dagger)\cr
&=
-\kappa+\partial\bar\partial\,\tr(CC^\dagger).\cr}}
Comparing this equation with \kalpot\ we get \tauK\ as was
to be shown.
This equation is very much reminiscent of the
Quillen holomorphic anomaly \ref\quil{D. Quillen,
Funkts. Anal. Prilozh. 19 (1985) 37.}.
This anomaly arises when one studies the determinant of
differential operators which depend holomorphically
on some parameters.  So naively the logarithm of the determinant is
expected to be a sum of a holomorphic and its conjugate piece.
There is an anomaly which causes a term involving both
holomorphic and anti-holomorphic  parameters to arise.
This mixing is computable, as is the
case with all anomalies, and in fact leads to recovering
the {\it full} determinant, i.e. even  the purely holomorphic
and anti-holomorphic pieces, by the requirement of
single-valuedness.  Our situation suggests
something similar should be happening here, as the mixing
of holomorphic and anti-holomorphic pieces are easy to compute
for the tau function \holq , and from that we can fix the other
terms by some global single-valuedness criteria.
The analogy may be in fact closer than it appears;
after all we {\it are} computing determinant
of a massive Dirac operator.  The fact that  there
is mass implies, however, that it is not a holomorphic operator.
But if the mass term is a `soft breaking' of the
holomorphicity, then we may expect that the holomorphic
anomaly should still be computable as we are finding.  It would be
interesting
to develop this line of thought further.

\newsec{The Tau Function as a Supersymmetric Index}

In the context of N=2 supersymmetry a physical quantity $A$ is called
an
{\it index} if it is invariant under arbitrary deformations of the
D-term. Such a quantity depends at most on $n$ parameters, where $n$
is
the dimension of $\CR$ (as the $F$-terms depend only on chiral
fields).

The $\tau$ function introduced in the previous sections is certainly
a
quantity of this kind. Indeed it can be computed from the
ground--state
metric $g$ which is itself invariant under deformations of the
D--term.
However the most interesting susy indices are those which can be
written
in the form ($\alpha$ is
a super--selection sector of the Hilbert space)
$$\Tr_\alpha\big[(-1)^F \CO e^{-\beta H}\big],$$
for some `natural' operator $\CO$. Only in this case they have a
simple path integral representation \pathrep.
For $\CO=1$ we get the original Witten index \witindex\
and for $\CO=F$ the index introduced in Ref.\newindex.

It is natural to ask whether the $\tau$ function can be written
in this way. In this section we discuss
some aspects of this issue.

\subsec{$\tau$ in the critical case}

We begin with the simpler case of a conformal family of
N=2 models.
In particular the couplings $t_i$
should correspond to exactly marginal
deformations.
In the conformal case it is more convenient to consider the K\"ahler
metric $K_{i\bar j}$ rather than the $\tau$ function itself.
The two are related by \tauK  .

In the critical case both right $F_R$ and left $F_L$ Fermi numbers
are
conserved.   Note that the first
try at the index, $Tr(-1)^F F_{L,R}q^{H_L}{\bar q}^{H_R}$ on
the {\it torus} vanishes by CPT. So, continuing the logic of
\newindex\
it makes sense to consider the following
object\foot{Conventions are as
follows. As usual, $q=\exp[2\pi i\varrho]$,  $\bar q=\exp[-2\pi i
\bar\varrho]$,
where $\varrho$ is the period of the
torus used to represent the functional trace
(we call it $\varrho$ to  avoid confusion with the $\tau$ function)
and
$\varrho_2={\rm Im}\, \varrho$. Moreover $2\pi H_L=L_0-c/24$ and
$2\pi H_R=\bar
L_0-c/24$.}
\eqn\anzt{\Tr\big[(-1)^F  F_R F_L\,   q^{H_L} \bar q^{H_R}\big].}
which is CPT even and generally non-zero\foot{It
is easy to show, using arguments similar to the
ones in the text that any pure insertion of the form
$F_L^n$ or $F_R^n$ is a `simple' index, in that it
is completely independent of target moduli.}.
Here the trace is over the periodic (Ramond) sector. We claim that
$K$ function defined
in previous section is given by
\eqn\taucri{K= -4\int_\CF {d^2\varrho\over \varrho_2}\,
\Tr\big[(-1)^F F_R F_L\, q^{H_L} \bar q^{H_R}\big]. }
where $\CF$ denotes the standard fundamental domain for
$SL(2,\BZ)$.  Notice that our definition of the $K$
in the previous section
fixes it up to addition of a holomorphic function plus its conjugate.
In many
interesting
cases the holomorphic piece can be determined uniquely by
exploiting the {\it target space} modular invariance \oneloop .
The above definition of $K$ has a fixed holomorphic piece
and since it depends only on the moduli of the target theory it
is automatically modular invariant.

Remarkably enough integrals similar to those in Eq.\taucri\
have already been studied by many
authors in a different context \oneloop, namely
in computing the
stringy one--loop correction to gauge
and gravitational couplings. The precise form above
has not been encountered.  However, it is plausible that
provided the central charge $\hat c$ has the correct value, \taucri\
gives the one--loop correction to the gravitational
coupling for a type II superstring `compactified' on the given N=2
superconformal model. In view of
 the geometric interpretation of $\log\tilde\tau$
as a K\"ahler potential for the
{\it torus} correlations, this is not too
surprising.

To show the above claim, let us vary the functional
representation of the RHS of \taucri\
with respect to the F--term couplings $t_i$, $\bar t_j$.
Notice
that no D--term perturbation is possible
 in the critical case. In fact these
perturbations are never marginal and so
will spoil conformal invariance which is
needed to  make sense out of
\taucri.
One obtains\foot{We have ignored regularization--dependent
contact terms. It is known
\oneloop\ that one can regularize \taucri\ in such a
way that these terms cancel while preserving conformal invariance.}
\eqn\dertra{\eqalign{\partial_i\partial_{\overline{j}}
&\Tr\big[(-1)^F  F_R F_L\, q^{H_L} \bar q^{H_R}\big]
=\cr
&=\varrho_2 \int\limits_{\rm torus} d^2z\,
\Tr\big[(-1)^F F_RF_L\, \{Q^+,[\bar Q^+,\bar
\phi_j(0)]\} \{\bar Q^-,[Q^-,\phi_i(z)]\} q^{H_L}\bar
q^{H_R}\big],\cr}   }
where we used translational invariance
to fix the position of $\bar\phi_j$ at the
origin extracting a factor of the area of the torus $\varrho_2$.
Also
the propagator $q^{H_L}\bar q^{H_R}$ is to be distributed among
the terms in the above expression corresponding
to where the point $z$ is inserted; but for simplicity
of notation we have written it as above and we continue
to use this notation throughout this section.
Now we use manipulations very similar to the ones
used in \newindex\ to simplify \dertra .  If  we take
$Q^+$ and take it around the trace it will cancel the
other term in the anti-commutator, except that we pick two
terms:
One coming from commutation with $F_L$ and the other one
by its anti-commutation with $Q^-$ which leads to $\partial \phi_i$
which vanishes upon integration (note that $Q^+$
commutes with $\phi_i$).  Similarly
taking $\bar Q^+$ around the trace the result
is non-vanishing only due to the
non-vanishing commutation of $\bar Q^+$ with $F_R$\foot{Indeed,
if $F_LF_R$ was not present our computation would
reduce to the usual argument
for the independence of the Witten index from any
deformation of the theory.} . Since
$$[F_L, Q^\pm]=\pm Q^\pm ,\qquad [F_R,\bar Q^\pm]=\pm \bar Q^\pm $$
So the integrand in \dertra\ takes the form
$$Tr\big[(-1)^F  \bar Q^+  Q^+ \bar \phi_j(0) \{\bar
Q^-,[Q^-,\phi_i(z)]\}
q^{H_L}\bar q^{H_R}\big]
$$
Now taking the $Q^-$ and $\bar Q^-$ around the trace the only
non-vanishing term comes from the anti-commutators
$$\{Q^-, Q^+\}=H_L \qquad \{\bar Q^-, \bar Q^+ \}=H_R$$
So finally one has
\eqn\dertraf{\partial_i\partial_{\overline{j}}
\Tr\big[(-1)^F F_R F_L
q^{H_L} \bar q^{H_R}\big] = \varrho_2\int\limits_{\rm
torus} d^2z\, \Tr\big[(-1)^F
 H_L\bar\phi_j(z) H_R\phi_i(0)   q^{H_L}\bar
q^{H_R}\big]. }
The integrand on the rhs of the above equation
looks like a total derivative in toroidal
moduli.  Indeed, consider the two--form
$$\Omega_{i\bar j}= i \tra{H_L
\int d^2z \ \bar\phi_j(z)H_R \ \phi_i(0) e^{i\varrho
H_L}  e^{-i\varrho^\ast H_R} }
d\varrho \wedge d\varrho^\ast,$$
then eq.\dertraf\ is rewritten as
\eqn\hhh{\partial_i\partial_{\overline{j}}K=-2
\int_\CF \Omega_{i\bar
j},}
and one can write $\Omega_{i\bar j}$ as a total derivative,
\eqn\diomega{\Omega_{i\bar j} = d\omega_{i\bar j},}
where
\eqn\defomega{\eqalign{\omega_{i\bar j} &
=\int_0^{\varrho_2} dt\, \tra{ \int
 d^2 z\  e^{Ht} \oint\bar \phi_j
e^{-Ht}H_R \phi_i e^{i\varrho H_L}
e^{-i\varrho^\ast H_R}} d\varrho^\ast\cr
&- \half \tra{\oint\bar\phi_j  e^{i\varrho H_L} e^{-i\varrho^\ast
H_R}\phi_i} d\varrho,\cr}  }
[as in sect. 3 we use the convention
$\oint \bar\phi_j(t)\equiv \int_0^1 dx\,
\bar\phi_j(x,t)$].
Thus the integral \hhh\ reduces to a boundary term
$$\partial_i\partial_{\overline{j}}
K= -2\int_{\partial\CF}
\omega_{i\bar j}.$$
A priori there are four
contributions to the LHS: the one coming from infinity,
those coming from the two vertical
segments $\varrho_1=\pm\half$ and finally that
of the arc $|\varrho|=1$.
However since $\omega_{i\bar j}$ is invariant under
$\varrho\rightarrow \varrho+1$, the contributions
from the two segments cancel each
other.  The contribution from the arc
cancels as well becuase the contribution of the
two half arcs cancel as they are mapped to one
another by the modular transformation $\rho\rightarrow
-1/\rho$.   Thus the only non-vanishing
contribution comes from the part of the boundary at infinity
(which is what one would expect since the moduli
space should be thought of as having only one boundary and that
is at infinity).

Let us compute this contribution.
One has
\eqn\boundary{\eqalign{&\left.- \int_0^{\varrho_2}dt\,
\tra{ \oint\bar\phi_j H e^{-Ht} \phi_i
e^{-H(\varrho_2-t)}}\right|_{\varrho_2\rightarrow\infty}+\cr
&+\left.\tra{\oint\bar\phi_j
e^{-H\varrho_2}\phi_i}\right|_{\varrho_2\rightarrow\infty},\cr} }
The first term in \boundary\ can be rewritten as
$$\eqalign{& \lim_{\varrho_2\rightarrow \infty}
  \int_0^{\varrho_2}
dt {d\phantom{a}\over dt} \tra{ \oint \bar\phi_j e^{-Ht} \phi_i P
}=\cr
&=-  \tra{\oint \bar\phi_j P^\prime \phi_i P },\cr}$$
where $P$ is the projector on the vacua and $P^\prime=
1-P$.
Now, since $\bar\phi_j$ and
$\phi_i$ {\it commute at equal time},
$$\tra{\bar\phi_j P^\prime \phi_i P}
=\tra{\bar\phi_j P \phi_i P^\prime}.$$

The second term in \boundary\ is
$$\lim_{\varrho_2\rightarrow \infty}
 \tra{\oint \bar\phi_j e^{-H\varrho_2}
\phi_i} =\tra{\oint \bar\phi_j P \phi_i }.$$
Then putting everything together,  the $\varrho_2=\infty$
contribution is
$$\eqalign{& -\tra{\oint\bar\phi_j P \phi_i (1-P)}+
\tra{\oint \bar\phi_j P \phi_i}=\cr
&= \tra{\oint \bar\phi_j P \phi_i P}=\cr
&= \tr[\bar C_j C_i],\cr}$$
which is the formula we wanted to show
$$\partial_i\partial_{\overline{j}}K=\tr[\bar C_j C_i].$$

Let us evaluate $K$ using this relation.
In the conformal
case the matrix $Q$ is just a constant (in an
appropriate basis). Then\foot{Notice that this formula differs for
a factor $2$ from what we would have guessed from the formula for the
massive case. This dicrepancy is related to subtleties in taking the
limit $\beta\rightarrow 0$.}
 $$\partial K=-\tr\big(Qg \partial g^{-1})=
\sum_k q_k \partial\log\big({\rm det}_k[g]\big),$$
where $q_k$ are the eigenvalues of $Q$,
and $\det_k[g]$ means the determinant of the
metric restricted to ground states
with $U(1)$ charge $q_k$.
In fact, using the constancy of $Q$, one has
$$\eqalign{\partial_{\bar j}\tr\big(Q g\partial_i g^{-1}\big)
&=\tr\big[Q \partial_{\bar j}(g\partial_i
g^{-1})\big]=\tr\big(Q[C_i,\bar
C_{\bar j}]\big)\cr
&=\tr\big([Q,C_i]\bar C_{\bar j}\big)=-\tr\big(C_i\bar C_{\bar
j}\big),\cr}$$
since, for a marginal deformation, $[Q,C_i]=-C_i$.
Then\foot{Consider a CY 3--fold.
Then,
this equation reads
$$K=-(m+3)\log\langle\bar 1|1\rangle- \log\det G +f +f^*,$$
where $G$ is the Zamolodchikov metric along the marginal (moduli)
directions and $m$ is the
number of moduli. Acting with $\partial_i\partial_{\bar j}$ on the
equation and recalling that $-\log\langle\bar 1|1\rangle$ is the
K\"ahler
potential for $G$, we get
$$\tr(C_i\bar C_j)=(m+3)G_{i\bar j}-R_{i\bar j}.$$
which gives a relation between the Ricci curvature of $G$ and the
ring coefficients. In fact, this relation is a well
known fact in special geometry (see \eg\ Eq.(38) of
\strominger ). The equation in the
text extends this identity to general critical models.}
\eqn\taucrit{e^{-K}= \prod_k \big({\rm det}_k[g]\big)^{-q_k}.
}
In particular, for the $\sigma$--model with target space a torus of
period $\varrho$, we have $\exp[-K]=({\rm Im}\, \varrho)^2$.
Since $K$ should depend only on the target space QFT, i.e.,
should be a modular invariant object, we can fix
the holomorphic plus the antiholomorphic piece of $K$, by
adding $2\rm log (\eta^2 \bar \eta^2)$ to the (logarithm of) tau
function.
This is the trick which is well known in the literature \oneloop .
This is an example of why the equation \tauK\
may be viewed as a {\it holomorphic anomaly} equation.

\subsec{The Massive Case}

Next we try to generalize the above discussion to the massive case.
In this case the two
chiral charges $F_L$, $F_R$ are not
conserved any longer. Only the vector
combination $F=F_L-F_R$ is. So  \anzt\
makes no sense. However this is not
really a problem. Going trough
 the computations for the critical case we see that we can
replace $F_LF_R$ by $-F^2/2$ and all
the arguments work as before\foot{Note that the difference between
these
two definitions at
the critical point is a term involving $F_L^2+F_R^2$ which
is completely independent of target moduli.
Or we can just take the integral with $F^2$ inserted over the doubled
fundamental domain which includes the image of the fundamental
domain under $\rho \rightarrow -1/\rho$ and use the fact
that under this modular transformation $F_L-F_R\rightarrow
i(F_L+F_R)$
to show that the additional pieces cancel out.}.

Then we consider the expression
\eqn\tramass{\Tr_\beta\big[(-1)^F F^2 e^{-t H}e^{i xP}\big],}
where the index $\beta$ means
that we quantize the theory on a circle of
perimeter $\beta$. Then \tramass\
can be written as a periodic path
integral over a torus of periods $(\beta, x+it)$.
We write $\varrho=(x+it)/\beta$
for the normalized period.
Then all manipulations above hold for the massive
case too. In particular
\eqn\dertram{i\partial_i\partial_{\overline{j}}
\Tr_\beta\big[(-1)^F F^2 e^{-t H}e^{i xP}\big]
{d\varrho\wedge d\varrho^\ast\over
\varrho_2} =d\omega_{i\bar j} }
where $\omega_{i\bar j}$ is as in \defomega.
So even in the massive case  the second
variation is a pure boundary term.
The contribution from $t\rightarrow
\infty$ is computed just
as in the critical case and we get
$$\beta^2\tr\big[C_i\bar C_j\big].$$

However there is a very important
difference with respect to the previous case. In
the massless theory we had modular
invariance and there was a natural integration
region, namely $\CF$. In the massive
case there is no natural integration
region.
If we choose to integrate over $\CF$,
the contributions from the two vertical
segments still cancel, since
the massive partition function
is invariant under $\varrho\rightarrow
 \varrho+1$ (as a consequence of
quantization of momentum in a periodic box).
But the arc contribution
{\it will not cancel}. Indeed the
argument for cancellation uses explicitly the
conformal invariance. A careful
evaluation of the arc contribution shows
that it is proportional
to\foot{Here $\langle\cdots\rangle_\varrho$ means the
path--integral over a
 torus whose (normalized) period is $\varrho$.}
$$\beta{\partial\phantom{\beta}\over
\partial\beta}\int\limits_{\pi/3}^{2\pi/3}d\theta\,
 \langle\phi_i(0)\int
d^2z\bar\phi_j(z)\rangle\Big|_{\varrho=e^{i\theta}},$$
i.e. to the variation of
$\langle\phi_i \int d^2z \bar \phi_j\rangle$ under an
infinitesimal rescaling.
This term vanishes only if the theory is conformal
and $\phi_i$, $\bar\phi_j$
are truly marginal deformations.
By the same argument,
the integral of \tramass\
over $\CF$ is not invariant under deformations of the
D--term. Indeed if we make
a change $\delta\kappa$ of the D--term we again get a
non--vanishing contribution from the arc proportional to
$$\beta{\partial\phantom{\beta}\over
\partial\beta}\int\limits_{\pi/3}^{2\pi/3}
d\theta\, \langle\delta
\kappa\rangle\Big|_{\varrho=e^{i\theta}}.$$

{}From a different point of view,
the problem can be seen as due to UV divergences.
Formally $\log\tau$ is given by
\eqn\formally{``\, \Tr[(-1)^F F^2 \delta(P) \log H]\, ".  }
One way to see this is to recall that in
the Ising case the $\tau$ function is the
determinant of the twisted Dirac operator.
Using the relation between the Dirac
operator and the supercharges
we get something like \formally.
One has to give a meaning to
this purely formal object. We can represent it
as the integral of \tramass\
over the strip $\varrho_2\geq 0$, $|\varrho_1|\leq
\half$. However this definition is
still badly divergent as $\varrho_2\rightarrow 0$.
In the critical case we know
that this divergence is fake, as it is due to  summing over
infinite copies of the fundamental domain.
To get a finite answer one has just
restrict to a single copy.
Unfortunately in the massive case this natural
regularization is not available.

However the UV divergent piece
of \dertram\ has also the form
$\partial_i\partial_{\bar j}F$
for some $F$. Then we can subtract $F$ from the
formal definition of  $\tau$,
getting something finite and with the
correct properties\foot{The same remark
applies to the regularization--dependent
contact terms.}.
The only drawback of this procedure, is
that in general it leads
to objects depending on the D-terms.
The simplest version of this subtraction is as
follows. In sect.3 we introduced a
family of K\"ahler potentials $K(\varrho,\beta)$.
Then the combination of K\"ahler potentials
$$K-\beta{\partial\phantom{\beta} \over
\partial\beta}\int\limits_{\pi/3}^{2\pi/3}
 d\theta\, K(e^{i\theta},\beta),$$
has no UV problem and it is in fact given
by the integral of \tramass\ over the
standard domain $\CF$. However this
combination is not really an index, since it
depends on the D-terms.

To get something more `canonical' one
 has to control the UV structure of the
theory in more detail.    There is a trick
which allows one to control the UV structure
whenever the
configuration space has   $\pi_1 =\bf Z$.
One takes the difference of
\tramass\ computed in two sectors of
 the Hilbert space carrying different
representations of the fundamental group.
The typical instance is the sine--Gordon model, i.e. the LG model
with
 superpotential $W(X)=\lambda\cos(X)$.
If one identifies $X\sim X+2\pi$, the
 configuration space becomes a circle and
$\pi_1=\bf Z$. Then we  introduce the topological
charge (=instanton number)
$$Q={1\over 2\pi}\int dt\, d_t X .$$
The trace over a $\theta$--sector of
 the Hilbert space are defined as
$$\Tr_\theta[(-1)^F \CO e^{-Ht}]
=\int[d\Phi] \CO \exp\{-S[\Phi]+i\theta Q\}.$$
The problem we had with integrating
over the full strip and getting infinity simply because
we are adding infinitely many
equivalent contributions does not exist anymore,
because once we use twisting with the $\pi_1$
the modular transforms are all inequivalent.
This is a well known trick, already used in the conformal
case in \ref\pol{J. Polchinski, Comm. Math. Phys. 104 (1986) 37.}.
Then we consider the integral over the full strip (and by
abbreviating
the integral and writing its relevant piece which
is the integral over $t=\rho_2$; the integral over $\rho_1$
simply projects to $P=0$ subsector) of the difference
\eqn\keidiff{K(\theta)-K(\theta^\prime)=2
\int_0^\infty{dt\over t}
\Big\{\Tr_\theta[(-1)^F F^2 e^{-Ht}]
-\Tr_{\theta^\prime}[(-1)^F F^2
e^{-Ht}]\Big\}.  }
This
difference is well defined (just as in \pol ).
Indeed, varying the D--term we get
$$\eqalign{\int_0^\infty dt {d^2\phantom{a}\over dt^2}&
\Big\{\Tr_\theta[(-1)^F
 \delta\kappa e^{-Ht}]-\Tr_{\theta^\prime}[(-1)^F
\delta\kappa e^{-Ht}]\Big\}\cr
&={d\phantom{a}\over dt}
\Big\{\Tr_\theta[(-1)^F
 \delta\kappa e^{-Ht}]-\Tr_{\theta^\prime}[(-1)^F
\delta\kappa e^{-Ht}]\Big\}
\bigg|_{t=0}^{t=\infty}.\cr}$$
The contribution from $t=\infty$
 vanishes since for large times the quantity in
brace reduces to a constant (up
to exponentially small terms). The contribution
from the boundary $t=0$ also vanishes.
 Indeed for any operator $\CO$
\eqn\bound{\Tr_\theta[(-1)^F
\CO e^{-Ht}]-\Tr_{\theta^\prime}[(-1)^F \CO
e^{-Ht}]= O(e^{-c/t})\qquad {\rm as}\ t\rightarrow 0.}
This equation also shows that the
 contribution from the boundary at 0 cancels
in $\partial_i\partial_{\bar j}
[K(\theta)-K(\theta^\prime)]$. Instead the
contribution from $t=\infty$ gives
$$\partial_i\partial_{\bar j}
[K(\theta)-K(\theta^\prime)]=\tr_\theta(C_i\bar
C_j)-\tr_{\theta^\prime}(C_i\bar C_j).$$
 To show \bound\ we use the path integral
representation of the RHS.
We write $\langle\cdots\rangle_\theta$ for
$\Tr_\theta[(-1)^F\dots e^{-Ht}]$.
One has $\langle\CO\rangle_\theta=\sum_n
e^{in\theta}\langle\CO\rangle_n$,
where $\langle\cdots\rangle_n$ denotes
path--integral in the $n$ instanton sector.
{}From this we see that the in difference
\bound\ only the non--trivial sectors $n\not=0$ contribute.
 Now for
small $t$
$$\langle \CO \rangle_n \sim e^{-(2\pi n)^2/t},$$
because  in this sector the action
is $\geq (2\pi n)^2/t$. This is easily seen
from the kinetic term
(the potential -- being positive
 definite --- cannot change
the conclusion) where we note that only the $P=0$ part
contributes
$$t\int_0^t ds d_s{\bar X} d_s X
 \geq \left|\int_0^t d_s X ds\right|^2=(2\pi
n)^2.$$
In eq.\keidiff\ $\theta$ labels
the regular solutions to PIII. We fix
$\theta^\prime$ to have the
value $\pi$ (corresponding to the trivial solution to
PIII $u=0$). Then
$$\log\tilde\tau(\theta)= K(\theta)-K(\pi),$$
corresponds to the standard definition
of the $\tau$ function for the regular
solution of PIII
 corresponding\foot{In fact we have
$$\partial_i\partial_{\bar j}K(\pi)=
\tr(C_iC_j^\dagger).$$} to $\theta$.
So we see that the term which
arises from the ``UV subtraction" is actually needed
to get the correct answer.

It would be very important to generalize this to the case
where there is no non-trivial $\pi_1$.  In the context of SQM
this may be possible to do by subtracting
from \tramass\ its asymptotic expansion
as $t\rightarrow 0$ (so that
only `exponentially small' terms remain).  This would
be interesting to study in more detail.

\subsec{ Relation with the
Ray--Singer Torsion}

The definition of the function $K$ \taucri\ is reminiscent of the
Ray--Singer analytic torsion \raysinger .  In particular,
in the context of supersymmetric $\sigma$--models,
interpreting $F_L$ and $F_R$ in terms of holomorphic
and anti-holomorphic degree of differential forms
and noting that the integral over moduli space is essentially what
is needed to give logarithm of the determinant of $H$ acting
on the differential forms allows us to
build up a dictionary between the two problems.
In fact more is true:  (A linear combination of) analytic torsion is
just
$\log\tilde\tau$ for a special class of
N=2 models.
Let us explain this. The $1d$ susy $\sigma$--models satisfy
the same \ttstar\ equations as
the $2d$ ones \topatop\ (although, in general,
with a different, much simpler, $\CR$ \ie\
the classical cohomology ring, without the instanton corrections
which makes the $2d$ case more interesting).

 We consider the
$1d$ $\sigma$--model with target space a compact
K\"ahler manifold  $\CM$ having a
non--trivial $\pi_1$.
It is a well known fact that the
Hilbert space of this model consists of all
(square--summable) $(p,q)$--forms
taking value in the flat bundles $V_\chi$
associated to unitary representations
$\chi$ of $\pi_1(\CM)$. The existence of
these different $\chi$ sectors comes from
the fact that various classes of maps
$S^1\rightarrow \CM$ can be weighted with
different phases. The susy charges act on
this space as $\bar\partial$, $\partial$
and their adjoints. Then the Hamiltonian is the usual Laplacian,
 \eqn\hamiltonian{\Delta= \bar\partial
\bar\partial^\dagger+\bar\partial^\dagger
\bar\partial, \qquad {\rm where}\quad
\bar\partial^\dagger=-\ast\partial\ast. }
 The susy vacua are precisely
the harmonic $(p,q)$--forms with coefficients
 in $V_\chi$. The  operators
$\phi_i$ are associated to the corresponding
cohomology classes, and act on the
Hilbert space by wedge product. Then $\CR$ is
 just the cohomology ring of $\CM$
(with coefficients on the flat bundles $V_\chi$).
In particular $\CR$ is
nilpotent and hence has the algebraic
structure typical of a critical
theory\foot{However, the \ttstar\ metric $g$ is
 a much simpler object in this
case. For forms $\Omega_j$ of degree $k\leq n$
having the form
$L^r v_j$ with $v_j$ primitive, the ground state metric  is
$$g_{j\bar l}=(-1)^r (-1)^{k(k-1)/2} {r!\over (n-k+r)!}\int
\omega(t,\bar t)^{n-k+2r}\wedge v_j\wedge C v_l^\ast$$
where $\omega(t,\bar t)$ is twice
the K\"ahler form (seen as a function of the couplings $t$),
and $C$ is the Weil operator acting on a $(p,q)$--forms as $i^{p-q}$.
This follows from \weilA\ \S I.4. The other
entries of $g$ can be obtained from \greality .
This $1d$ result serves also as
a boundary condition for the $2d$ \ttstar\
equations, see \sigmamodels. }. The
conserved charges $F$ and $Q^5$ act on
$(p,q)$ forms as $(p-q)$ and $(p+q-n)$,
respectively. The topological metric
$\eta_{ij}$ is just the intersection form
in cohomology \TFT.  Comparing with the
Hodge metric, we see that the
real structure acts on forms as
$\alpha^\star=\ast\alpha^\ast$. Then, if
$\Omega_k$ is a `canonical' basis
of harmonic forms, we have
\eqn\Oreality{\ast\Omega_k^\ast= g_{l\bar k}\Omega_l,  }
since, the real structure acts on
the vacua  as the matrix $g\eta^{-1}$ \greality.

We consider a family of such models of the
form
$$S=S_0+\left(\sum_k t^k\int ds \{Q^-,[\bar Q^-,\widehat\omega_k]\}
+{\rm c.c.}\right)$$
where $\widehat\omega_k\equiv
(\omega_k)_{i\bar j}\psi^i\psi^{\bar j}$ and the
$\omega_k$'s give a basis of $H^{1,1}(\CM)$.

For each $p=1,\dots, n={\rm dim}\, \CM$,
the $\bar\partial$--torsion $T_p(\chi)$is
defined as \raysinger\
\eqn\anatorsion{\eqalign{
&\log T_p(\chi)=\half \sum_{q=0}^n (-1)^q\,
{d\phantom{s}\over ds}\zeta_{pq}(s,\chi)\Big|_{s=0},\cr
&\zeta_{pq}(s,\chi)={1\over \Gamma(s)}\int_0^\infty x^{s-1} dx
\, \Tr_{(p,q)} \Big[\CP_\chi\left(e^{-Hx}-\Pi\right)\Big].\cr}   }
Here $Tr_{(p,q)}$ is the trace over the
$(p,q)$ sector of the Hilbert space,
 $\CP_\chi$ is the projector on the $\chi$
representation, and $\Pi$ is the
projector on the ground states. The basic
mathematical fact is that the {\it
difference} of  the torsions for two $\chi$'s depends only on the
{\it
class} of the K\"ahler metric.
In physical language this is independence from
the D--term. Then this difference
is a susy index in the sense of \newindex .

The relation with the N=2 $\tau$ function
 is as follows. Let $\chi$ and
$\chi^\prime$ be real representations. Then
$$2\sum_p(-1)^p p\Big[\log T_p(\chi)
-\log T_p(\chi^\prime)\Big]=\sum(-1)^{p+q}pq\zeta'_{pq}(0,\chi
)-(\chi
\rightarrow \chi ')$$
\eqn\tauRS{=
-\log\tilde\tau(\chi)+\log\tau(\chi^\prime).  }
In fact,
let $\CA_i$ be the operator which acts on
forms as $\ast^{-1}(\partial_i\ast)$.
Then from \hamiltonian\ one gets
$$\partial_i\left\{\sum_q (-1)^q q \,
\Tr_{(p,q)}\Big[\CP_\chi\left(e^{-Hx}
-\Pi\right)\Big]\right\}=\sum_q (-1)^q
\, {d\phantom{x}\over
dx}\Tr_{(p,q)}\Big[\CP_\chi\left(\CA_ie^{-Hx}\right)\Big].$$
This equation together with \anatorsion\ give \raysinger\
$$\eqalign{&\partial_i \sum_p(-1)^p p\Big[\log T_p(\chi)-\log
T_p(\chi^\prime)\Big]=\cr
& =\half\sum_p(-1)^{p+q} p \int_0^\infty dx\,
{d\phantom{x}\over dx}\left\{\Tr_{(p,q)}\Big[\CP_\chi(\CA_i
e^{-Hx})\Big] -\Tr_{(p,q)}
\Big[\CP_{\chi^\prime}(\CA_i e^{-Hx})\Big]\right\}\cr
&=\half \sum_p (-1)^{p+q} p \,
\left\{\Tr_{(p,q)}\Big[\CP_{\chi}(\CA_i\Pi)\Big]
-\Tr_{(p,q)}\Big[\CP_{\chi^\prime}(\CA_i \Pi)\Big]\right\},\cr}$$
where we used that, as $x\rightarrow 0$,
\eqn\expsmall{\Tr_{(p,q)}\Big[\CP_\chi(\CA_i\,
e^{-Hx})\Big]-\Tr_{(p,q)}\Big[\CP_{\chi^\prime}(\CA_i\,
e^{-Hx})\Big]=O(e^{-c/x}).  }
{}From \Oreality\ it follows that
$$\CA_i\, \Pi = -(-1)^F g\partial_i g^{-1},$$
and hence we have
$$\eqalign{ \sum_p &  (-1)^{p+q} p\,
\Tr_{(p,q)}\Big[\CP_{\chi}(\CA_i\Pi)\Big]
=\cr &=-\half \tr(\CP_\chi Q
g\partial_i g^{-1})-\half
 \tr(\CP_\chi F g\partial_i g^{-1})\cr
&= -\partial_i\log\tilde\tau(\chi)-\half
 \tr(\CP_\chi F g\partial_i
g^{-1}).\cr}$$
The additional term in the rhs was not
present in our definition of the
$\tau$ function just because in
Ising--like models all vacua have $F=0$; this
term should be added to the definition
of the $\tau$--function for the general
case. Anyhow,  for $\chi$ real this
term vanishes by PCT, and we get Eq.\tauRS.

To construct a general theory of susy
$\tau$--functions we have to extend
this  heat kernel argument in two directions: {\it i)} to more
general rings than those arising from cohomology; and {\it ii)}
to  $2d$ field theory, \ie\ to loop spaces.  This discussion
suggests that the $\tau$--function which is defined
and is computable for
an arbitrary $2d$ susy $\sigma$--model (at least with $c_1>0$),
is the generalization of the analytic torsion to the loop space
of K\"ahler manifolds.  This is an exciting
mathematical direction to pursue further.

\newsec{Conclusion}

We have seen that the Ising model can be viewed as
a `target space' description for $N=2$ QFT's in two dimensions.
The equations describing the geometry of $N=2$ ground
states (the $t t^*$ equations) are the same equations
which characterize the spin correlations for the massive
Ising model.
In particular the tau function for the Ising model
is a new supersymmetric index for $N=2$ theories.  As we have
seen this index is related to a generalization of
Ray-Singer analytic torsion to the loop space of K\"ahler manifolds.
This index is essentially fixed by its holomorphic anomaly,
which can be computed in terms of the metric on the ground
states and the chiral ring.
Moreover, the tau function can also be interpreted as a `canonical'
K\"ahler
potential for the moduli space of $N=2$ theories which leads to a
K\"ahler
metric on this space.

There are many directions worth pursuing.
Probably the most important one is to tame
the ultraviolet divergencies in the path-integral
formulation for index for the generic massive theory.
We were able to do this only for the conformal case
(where the fundamental region automatically
cuts off UV divergencies) and for massive cases with
non-trivial $\pi_1$ for the configuration space.

Another direction which would be interesting to explore
is to find the corresponding `target theory' for other
$N=2$ theories.  Since in some special cases we
got the interesting model of massive Ising model,
it may be that for other cases we may also get
interesting 2d QFT's (perhaps with fields having non-abelian braiding
properties).

Another possible application of these ideas is to $N=2$ strings
(see \ref\ov{H. Ooguri and C. Vafa, Nucl. Phys. B361 (1991) 469;
Nucl. Phys. B367 (1991) 83.} and references in it).  In this
case if the target space is the cotangent of the torus,
the one-loop partition function is the same as the tau function
defined here!  It would be interesting to see if this is true
for more general compactifications.  Indeed it
was conjectured in \ov\ that the partition function may be
characterized by a holomorphic anomaly condition, which is
what we have found here to be the case for the tau function.

We would like to thank B. Dubrovin, E. Gava, A. Lesniewski,
G. Mussardo, K.S. Narain and I. Singer for valuable discussions.
C.V. would also like to thank ICTP for its kind
hospitality where part of this work was done.

The research of C.V. was supported in part by Packard fellowship
and NSF grants PHY-87-14654 and PHY-89-57162.

\listrefs
\end